\documentclass[12pt]{article}

\usepackage{jheppub}
\usepackage{amsmath,amssymb,euscript,array,mathrsfs}

\usepackage{epsfig}

\def\XXint#1#2#3{{\setbox0=\hbox{$#1{#2#3}{\int}$}
     \vcenter{\hbox{$#2#3$}}\kern-.5\wd0}}

\def\a{\alpha}
\def\b{\beta}
\def\c{\gamma}
\def\d{\delta}
\def\e{\epsilon}

\def\l{\lambda}
\def\m{\mu}
\def\n{\nu}

\def\r{\rho}
\def\s{\sigma}
\def\t{\tau}
\def\u{\upsilon}
\def\w{\omega}

\def\D{\Delta}
\def\L{\Lambda}

\def\BC{{\boldsymbol C}}

\def\Bh{{\boldsymbol h}}

\def\Bk{{\boldsymbol k}}

\def\tr{{\rm tr}}

\def\det{{\rm det}}

\def\BDelta{\boldsymbol{\Delta}}

\def\BA{\boldsymbol{A}}
\def\BB{\boldsymbol{B}}
\def\BC{\boldsymbol{C}}

\def\BOmega{\boldsymbol{\Omega}}
\def\bomega{\boldsymbol{\omega}}
\def\bepsilon{\boldsymbol{\epsilon}}

\def\Dbarslash{\,\,{\raise.15ex\hbox{/}\mkern-12mu {\bar D}}}
\def\Dslash{\,\,{\raise.15ex\hbox{/}\mkern-12mu D}}
\def\delslash{\,\,{\raise.15ex\hbox{/}\mkern-9mu \partial}}
\def\delbarslash{\,\,{\raise.15ex\hbox{/}\mkern-9mu {\bar\partial}}}

\def\F{{\mathfrak F}}
\def\G{{\mathfrak G}}

\def\rta{\rightarrow}
\def\tr{{\rm tr}}
\def\dprime{{\prime\prime}}

\title{\begin{center}A New Twist on the Geometry of \\
Gravitational Plane Waves  \\
\end{center}}
\author{Graham M. Shore}

\affiliation{Department of Physics,\\ 
College of Science,\\
Swansea University,\\
Swansea,\\ 
SA2 8PP, UK.}

\emailAdd{g.m.shore@swansea.ac.uk}

\abstract{The geometry of twisted null geodesic congruences in gravitational plane wave 
spacetimes is explored, with special focus on homogeneous plane waves. 
The r\^ole of twist in the relation of the Rosen coordinates adapted
to a null congruence with the fundamental Brinkmann coordinates is explained and a 
generalised form of the Rosen metric describing a gravitational plane wave is derived.
The Killing vectors and isometry algebra of homogeneous plane waves (HPWs) are described in both 
Brinkmann and twisted Rosen form and used to demonstrate the coset space structure of HPWs.
The van Vleck-Morette determinant for twisted congruences is evaluated in both Brinkmann 
and Rosen descriptions. The twisted null congruences of the Ozsv{\'a}th-Sch{\"u}cking,
`anti-Mach' plane wave are investigated in detail.
These developments provide the necessary geometric toolkit for future investigations of the 
r\^ole of twist in loop effects in quantum field theory in curved spacetime, where gravitational 
plane waves arise generically as Penrose limits; in string theory, where they
are important as string backgrounds; and potentially in the detection of 
gravitational waves in astronomy.

\vskip1cm}

\begin{document}

\maketitle

\setlength{\parskip}{10pt}


\section{Introduction}\label{sect 1}

Gravitational plane waves arise generically as the Penrose limits 
\cite{Penrose:1965rx, Penrose} of an arbitrary curved spacetime 
and an associated null geodesic $\gamma$. They are a truncation of the full background spacetime
which captures the essential geometry of geodesic deviation around $\gamma$. 
Expressed in Brinkmann coordinates,
the metric is 
\begin{equation}
ds^2 = 2 du dv + h_{ij}(u) x^i x^j du^2 + (dx^i)^2 \ .
\label{a1}
\end{equation}
These Brinkmann coordinates are identified as Fermi normal coordinates associated with the 
null geodesic $\gamma$ in the original spacetime \cite{Blau:2006ar},
while the profile function $h_{ij}(u)$ is given by the curvature component $R_{uiuj}$ 
which appears in the Jacobi equation describing geodesic deviation.

This is precisely the feature of the geometry required to describe vacuum polarisation,
or loop, effects on the propagation of quantum fields in curved spacetime
\cite{Hollowood:2007ku, Hollowood:2008kq, Hollowood:2009qz, Hollowood:2011yh}.
This is evident in a worldline formalism, or equivalently using the Schwinger-de Witt
representation, where the propagator is expressed in terms of classical trajectories.
Loop corrections are determined by the geometry of the set of geodesics neighbouring
the classical path -- that is, by the geodesic congruence in which the classical
null geodesic is embedded.

In previous work, quantum loop effects have been shown to lead to a variety of novel
effects related to the induced violation of the strong equivalence principle, including
ultra-high energy scattering \cite{Hollowood:2015elj, Hollowood:2016ryc}
and the generation of matter-antimatter asymmetry 
\cite{McDonald:2015ooa, McDonald:2015iwt, McDonald:2016ehm},
challenging conventional understanding of how fundamental concepts such as causality,
analyticity and unitarity are realised in quantum field theory in curved spacetime
\cite{Hollowood:2007ku, Hollowood:2008kq, Hollowood:2009qz, Hollowood:2011yh}.

The propagation of quantum fields in a general curved spacetime background 
is therefore determined by the geometry of (null) geodesic congruences in the 
corresponding gravitational plane wave realised in the Penrose limit.
This is the central motivation for the study of plane wave geometry and
null congruences presented here, although the novel geometry we will describe
is in principle of much wider relevance. As discussed briefly in section \ref{sect 8},
this could also include potential applications to the detection of astronomical 
gravitational waves. 

Conventionally, studies of the propagation of, for example, electromagnetic fields in curved
spacetime have centred on electromagnetic plane waves, in a geometric optics description
where the tangent vector fields characterising the null rays are given as the derivative of the
plane-wave phase, $k_\m = \partial_\m \Theta$, the rays being normal to surfaces of
constant phase. Such `hypersurface-forming' rays are therefore by definition
a gradient, rotationless, flow. However, in general, geodesic congruences are described
by the Raychoudhuri equations in terms of three optical scalars $\theta$, $\s$ and $\omega$,
representing expansion, shear, and rotation or `twist'. 
Clearly the twist vanishes for a gradient flow.

In this paper, we establish the mathematical framework to allow a more general analysis
of quantum field propagation in curved spacetime by developing the geometry of
{\it twisted} null congruences in gravitational plane wave backgrounds.   

An important element is the relation of Brinkmann and Rosen coordinates. 
Whereas the original 
Brinkmann coordinates for the plane wave metric (\ref{a1}) are more fundamental, the Rosen
description is tailored to a choice of geodesic congruence. This explains why many
calculations of loop effects in QFT are most simply and elegantly performed in Rosen
coordinates. Here, we introduce the Rosen description appropriate to a twisted null
congruence, in which case the metric takes the novel form,
\begin{equation}
ds^2 = 2 du dV - 2 X^a {\bomega}_{ab} dX^b du + dX^a {\BC}_{ab}(u) dX^b \ .
\label{a2}
\end{equation}
Here $\BC_{ab} = \left(E^T E\right)_{ab}$, where $E^i{}_a(u)$ is the zweibein relating the 
transverse Brinkmann and Rosen coordinates, $x^i = E^i{}_a(u) X^a$, and the 
twist enters through $\bomega = E^t \omega E$.
The new feature is of course the off-diagonal term involving the twist.

The origin of twist and the subtleties of the Brinkmann-Rosen relation are first explained,
then we study the Rosen metric (\ref{a2}) in detail in its own right, finding the geodesics,
wave equation solutions, the nature of the isometries and the explicit form of the
van Vleck-Morette matrix.

The van Vleck-Morette (VVM) determinant, or more generally the VVM matrix, is perhaps 
the most important geometric quantity influencing quantum field propagation in curved 
spacetime. It enters the Schwinger-de Witt representation of the propagator
and all loop effects, including modified dispersion relations, are formally expressed 
in terms of it \cite{Hollowood:2007ku, Hollowood:2008kq, Hollowood:2009qz, Hollowood:2011yh}.  
Since it is essentially linked to the geodesic congruence, its evaluation 
is fundamentally altered in the presence of twist. Here, we derive the general form of the 
VVM matrix related to twisted null congruences, showing the equivalence of the Brinkmann 
and Rosen descriptions. 

Another important theme of the paper is the analysis of isometries. We first give an account of the 
isometry algebra, and its extension for the homogeneous plane waves discussed below,
using the Brinkmann metric (\ref{a1}), and show how the gravitational plane wave
spacetime may be represented as a coset space \cite{Papadopoulos:2002bg, Blau:2003rt}. 
Then, we reconsider the isometries 
from the point of view of the Rosen metric (\ref{a2}) and show how the form of the corresponding 
Killing vectors and commutation relations depend on the twist, generalising the analyses of
\cite{Blau:2002js, Duval:2017els,Zhang:2017rno,Zhang:2017geq} for the twist-free case.

A particularly important class of gravitational plane waves are the {\it homogeneous plane waves} (HPWs). 
These are of special interest from several points of view, including their r\^ole as exactly solvable, 
time-dependent string backgrounds (see {\it e.g.}~\cite{Blau:2001ne, Papadopoulos:2002bg,
Blau:2002js, Blau:2003rt, Blau:2007db}).
Our discussion of the geometry of twist in these backgrounds may therefore be of relevance
also in string theory.
They occur importantly as Penrose limits -- in particular, HPWs of type II
(see below) arise in the near-singularity region of a broad class of black hole and cosmological 
spacetimes \cite{Blau:2003dz, Blau:2004yi}, \cite{Hollowood:2009qz}.

Homogeneous plane waves (HPWs) have been classified by Blau and O'Loughlin \cite{Blau:2002js} in a paper
which is key to the present work. They fall into two types, generalising the symmetric plane waves 
(Cahen-Wallach spacetime \cite{CW}) and singular homogeneous plane waves, respectively.
The corresponding metrics are, in Brinkmann coordinates,
\begin{equation}
ds^2 = 2 du dv +  \left(e^{\e u} \,h_0 \,e^{-\e u}\right)_{ij} x^i x^j\, du^2 + \left(dx^i\right)^2 \ ,
~~~~~~~~~~ ({\rm type~I}) ~~~~~~
\label{a3}
\end{equation}
and
\begin{equation}
ds^2 = 2 du dv + \left( e^{\e \log u} \, \frac{h_0}{u^2} \, e^{-\e \log u}\right)_{ij} x^i x^j \,du^2  
+ \left(dx^i\right)^2 \ ,
~~~~~~~~~~~({\rm type~II}) 
\label{a4}
\end{equation}
for a constant $2\times 2$ matrix $(h_0)_{ij}$, where $\epsilon_{ij}$ is the usual antisymmetric symbol.
Compared to the general plane wave metric (\ref{a1}), whose isometries satisfy a Heisenberg algebra
(see section \ref{sect 3}), these HPWs are characterised by an additional symmetry 
related to translations in the lightcone coordinate $u$. This extra degree of symmetry underlies 
their importance in many applications, especially in string theory.

In this paper, we focus on the geometry of twisted null congruences and the Rosen description for
metrics of type I, though our methods and conclusions will carry over more or less directly to type II
metrics as well.   

A special place in classical general relativity belongs to the {\it Ricci-flat} HPW metric of type I.
This is known as Ozsv{\'a}th-Sch{\"u}cking spacetime \cite{OS}, originally presented in the
literature in the form (our notation, see appendix \ref{sect 4.1})
\begin{equation}
ds^2 = 2 dU dW -2\sqrt{2} \,z^1 dz^2 \,dU + (z^1)^2\, dU^2 + \left(dz^i\right)^2  \ .
\label{a5}
\end{equation}
Clearly, while making obvious three of the isometries (translations in $U$, $W$ and $z^2$) associated
with the extended Heisenberg algebra, this choice of coordinates disguises the fact that this
is a gravitational radiation metric. This becomes manifest only after evaluating the Newman-Penrose 
curvature scalars and noting that it is indeed a Petrov type N spacetime.
Its interest in general relativity is due to the fact that it is a vacuum (Ricci-flat) solution
of Einstein's field equations which is geodesically complete and singularity free, yet is non-trivial in the
sense that it is not simply flat, Minkowski spacetime. In this sense, it violates Mach's principle
and we follow \cite{OS,Blau:2002js} in referring to it as an ``anti-Mach'' spacetime.

As we shall see, this type of spacetime very naturally contains geodesics which fall into twisted
null congruences, and we study their geometry in considerable detail in appendix \ref{sect 4}. 
The property of twist automatically leads to null geodesics which are periodic in $u$, and this
underlies the suggestion 
that the metric (\ref{a5}) allows the existence of closed null geodesics \cite{Sarma:2012jn}
(see also \cite{Sarma:2013zza}).  Our analysis does not confirm this, however, and we find, 
as expected on general grounds \cite{Marolf:2002ye},
that these type N homogeneous plane wave spacetimes are indeed causal.

The paper is organised as follows. In section \ref{sect 2}, we introduce the essential geometry and 
geodesics for a general gravitational plane wave metric in Brinkmann form, explain the origin of twist,
and derive the new Rosen metric (\ref{a2}). Homogeneous plane waves are introduced in section \ref{sect 3}
and their isometries and coset structure are explored. 
We return to the twisted Rosen metric in section \ref{sect 5} and, taking it on its own merits, discuss the
geodesic equations and solutions of the wave equation in this background. The Rosen isometries
are described at length from different points of view in section \ref{sect 6}. 
Section \ref{sect 7} contains the evaluation of the 
VVM matrix in both Brinkmann and twisted Rosen coordinates and the demonstration of their equivalence.
Some concluding remarks are given in section \ref{sect 8}.
Finally, in an extensive appendix, we give a detailed account
of twisted null congruences in the the generalised Ozsv{\'a}th-Sch{\"u}cking model, 
introducing co-rotating coordinates and a Newman-Penrose basis adapted to the null geodesics.

\section{Plane Waves, Twisted Null Congruences and Optical Scalars}\label{sect 2}

We begin by studying the geometry of a general gravitational plane wave and its null geodesics in the 
Brinkmann description, extending the conventional description to allow the possibility of
twisted congruences.\footnote{
There is an extensive literature in general relativity on gravitational plane waves and their 
symmetries and geodesic structure.
Some classic references on exact solutions of Einstein's equations including gravitational plane
waves are \cite{Stephani:2003tm, Griffiths:2009dfa,Griffiths:1991zp}, while a comprehensive
set of references may also be found in \cite{Zhang:2017geq}. 
See also \cite{Gibbons:1975jb} for an early discussion of quantum field theory effects and
\cite{Harte:2012jg,Harte:2015ila} for recent studies of
null geodesics and gravitational lensing in gravitational wave backgrounds. } 
The optical scalars and their Raychoudhuri equations are introduced,
and the metric is rewitten in a novel Rosen form appropriate to twisted null congruences.

\subsection{Brinkmann coordinates}\label{sect2.1}

The starting point is the gravitational plane wave metric in four dimensions in Brinkmann coordinates
$x^\m = (u,v,x^i)$,
\begin{equation}
ds^2 = 2 du dv + h_{ij}(u) x^i x^j du^2 + (dx^i)^2 \ ,
\label{b1}
\end{equation}
where the Riemann curvature tensor $R_{uiuj} = -h_{ij}$ is determined directly by the profile function.
The corresponding null geodesic equations are,
\begin{align}
\ddot{x}^i - h^i{}_j x^j &= 0 \ , \nonumber \\
\ddot{v} + \tfrac{1}{2} \dot{h}_{ij} x^i x^j + 2 h_{ij} \dot{x}^i x^j &= 0 \ . 
\label{b2}
\end{align}
Here, we have immediately taken $u$ as the affine parameter, the overdot denotes differentiation 
with respect to $u$,
and the transverse indices are $i = 1,2$.  The solutions to (\ref{b2}) can be written in terms of a zweibein
$E^i{}_a(u)$, with $a=1,2$, as follows:
\begin{align}
x^i &= E^i{}_a(u) X^a \ , \nonumber\\
v &= V - \tfrac{1}{2} {\bf \Omega}_{ab}(u) X^a X^b \ , 
\label{b3}
\end{align}
where we define the key quantity $\Omega_{ij}(u)$ by
\begin{equation}
\Omega_{ij} = \left(\dot{E} E^{-1}\right)_{ij} \ ,
\label{b4}
\end{equation}
and ${\bf \Omega}_{ab} = \left(E^T \Omega E\right)_{ab}$.    
The zweibein $E^i{}_a$ and $\Omega_{ij}$ are related to the profile function $h_{ij}$ in the metric
by
\begin{equation}
h_{ij} = \dot{\Omega}_{ij} + \left(\Omega^2\right)_{ij} \ .
\label{b5}
\end{equation}
Notice that nowhere in this construction have we assumed that $\Omega_{ij}$ is symmetric.

Now consider the null congruence of geodesics labelled by $X^a, V$, centred on a reference geodesic
$\gamma$, with tangent vectors $k^\m = d{x}^\m/du$.  Evaluating the first integrals of the 
geodesic equations, we find
\begin{equation}
k^\m = \dot{x}^\m = \begin{pmatrix} 
1 \\
-\tfrac{1}{2} x \left(\dot{\Omega} + \Omega^2 + \Omega^T \Omega\right) x \\
(\Omega x)^i\\ \end{pmatrix}
\label{b6}
\end{equation}
Now let $z^\m$ be the connecting vector between $\gamma$ and other elements of the congruence.
By definition, the Lie derivative of $z^\m$ along $\gamma$ vanishes, {\it i.e.}
\begin{equation}
{\mathcal L}_k z^\m = k^\n D_\n z^\m - \left(D_\n k^\m\right) z^\n \ ,
\label{b7}
\end{equation}
where $D_\m$ is the ordinary covariant derivative, and so
\begin{equation}
k^\n D_\n z^\m  = \Omega^\m{}_\n z^\n \ ,
\label{b8}
\end{equation}
where we define\footnote{Although at this point we only require the transverse projection
of $\Omega_{\m\n}$, we will need the full component form later. This is readily evaluated
using the covariant derivatives, with Christoffel symbols
\begin{equation*}
\Gamma^v_{uu} = \tfrac{1}{2}\dot{h}_{ij} x^i x^j \ , ~~~~~~~~
\Gamma^v_{ui} = h_{ij} x^j \ , ~~~~~~~~
\Gamma^i_{uu} = - h^i{}_j x^j \ ,
\end{equation*}
and we find
\begin{equation*}
\Omega_{\m\n} = \begin{pmatrix}
\dot{x} \Omega \dot{x} &0 & - (\dot{x}\Omega)_j  \\
0 & 0 &{\bf 0} \\
- (\Omega \dot{x})_i &{\bf 0} &\Omega_{ij} \\
\end{pmatrix}
\end{equation*}
where $\dot{x} = \Omega x$. 
Note that this is not symmetric unless $\Omega = \Omega^T$. \label{fn1}}
\begin{equation}
\Omega_{\m\n} = D_\n k_\m \ .
\label{b9}
\end{equation}
It is then clear that with this definition, $\Omega_{ij} \equiv D_j k_i$ coincides with the 
form (\ref{b4}) above. This shows very directly how $\Omega_{ij}$ determines how the transverse connecting
vector is parallel transported along the reference geodesic and therefore characterises the null congruence.
Note that the order of indices is important in (\ref{b9}) since $\Omega_{ij}$ is
not necessarily symmetric.

The optical scalars for this congruence are defined from $\Omega_{ij}$ as
\begin{equation}
\Omega_{ij} = \tfrac{1}{2} \theta \d_{ij} + \s_{ij} + \w_{ij} \ ,
\label{b10}
\end{equation}
where $\theta$ is the expansion, the symmetric trace-free tensor $\s_{ij}$ is the shear
and the antisymmetric tensor $\w_{ij}$ is the rotation or twist.

Note immediately that in the case where $k^\m$ represents a gradient flow, {\it i.e.} 
$k_\m = \partial_\m \Theta$, then $\Omega_{ij}$ is symmetric and the congruence has vanishing twist. 
Such a congruence is then said to be ``hypersurface-forming''. It is realised in the familiar case
where we consider $k_\m$ to be the wave vector of, say, an electromagnetic wave propagating in the 
spacetime (\ref{b1}), in which case $\Theta$ represents the phase; $k_\m$ is then normal to the surface of
constant phase.  Here, we relax this condition and allow $k_\m$ to be a general vector field.
In this more general situation, $\Omega_{ij}$ need not be symmetric and the null congruence 
may have a non-vanishing twist, $\w_{ij} \neq 0$.

The variation of the optical scalars along the congruence is described by the Raychaudhuri equations.
In fact, these can be obtained simply as a rewriting of (\ref{b5}) as                     
\begin{equation}
\dot{\Omega}_{ij} = h_{ij} - \left(\Omega^2\right)_{ij} \ .
\label{b11}
\end{equation}
Substituting (\ref{b10}), we find after a short calculation the individual variations for the 
optical scalars themselves, {\it viz.}
\begin{align}
\dot{\theta} &= - \tfrac{1}{2} \theta^2 - \tr\, \s^2 - \tr\, \w^2 - R_{uu} \ , \nonumber\\
\dot{\s}_{ij} &= - \theta \s_{ij} - C_{uiuj} \ , \nonumber \\
\dot{\w}_{ij} &= - \theta \w_{ij} \ ,
\label{b12}
\end{align}
where $C_{uiuj}$ is the Weyl tensor.  Note also the useful formula 
\begin{equation}
\dot{\w} = - \Omega \w - \w \Omega^T \ ,
\label{b122}
\end{equation}
which follows directly from (\ref{b11}) with the condition that $h_{ij}$ is symmetric.

\subsection{Rosen coordinates}

An important alternative description of the gravitational plane wave is in terms of Rosen coordinates.
Although the Rosen form of the metric in general has unphysical coordinate singularities, it is 
adapted to the nature of the geodesic congruence we wish to consider and so plays a key role
in our discussion. Clearly, since it is tied to a particular choice of congruence, the Rosen metric is
not unique.

To make this coordinate transformation, we take the $X^a, V$ from the geodesic solutions (\ref{b3})
and define Rosen coordinates $(u, V, X^a)$ as
\begin{align}
X^a &= \left(E^{-1}\right)^a{}_i x^i \ , \nonumber \\
V &= v + \tfrac{1}{2} x^i \Omega_{ij} x^j \ .
\label{b15}
\end{align}
Expressing the Brinkmann metric (\ref{b1}) in terms of these coordinates, we find after a short 
calculation,
\begin{align}
ds^2 &= 2 du dV - X\left(\dot{\bf \Omega} - E^T h E - \dot{E}^T \dot{E} \right) X  du^2 \nonumber \\
&~~~- \left[dX \left({\bf \Omega} - E^T \dot{E}\right) X + X\left({\bf \Omega} - \dot{E}^T E\right) dX \right] du 
+ dX E^T E dX \ .
\label{b16}
\end{align}
Then, since from its definition as ${\bf\Omega}_{ab} = \left(E^T \Omega E\right)_{ab}$,\footnote{As far as 
possible, we use boldface notation for quantities such as ${\BOmega}$, ${\bomega}$, ${\BC}$ etc.~in 
Rosen coordinates related in this way to the fundamental definitions in Brinkmann coordinates.} we have
\begin{equation}
\dot{\bf\Omega} = \dot{E}^T \dot{E} + E^T \Omega^2 E + E^T \dot{\Omega}E \ ,
\label{b17}
\end{equation}
and using (\ref{b5}) for $h_{ij}$, we see that the coefficient of $du^2$ vanishes as in the conventional construction.
As usual, we also identify ${\BC}_{ab} = \left(E^T E\right)_{ab}$ as the transverse Rosen metric.
The novelty comes in the remaining term, where we must distinguish ${\bf\Omega} = E^T \dot{E}$ from its transpose
${\bf\Omega}^T = \dot{E}^T E$ when the zweibein is describing a congruence with twist.
Defining ${\boldsymbol\omega}_{ab} = \left(E^T \omega E\right)_{ab}$, we therefore find the Rosen metric
\begin{equation}
ds^2 = 2 du dV - 2 X^a {\boldsymbol\omega}_{ab} dX^b du + dX^a {\BC}_{ab}(u) dX^b \ .
\label{b18}
\end{equation}

This differs from the usual form of the Rosen metric for a plane wave spacetime due to the rotation term
involving the twist ${\boldsymbol\omega}$. Notice immediately that its equivalence to the 
Brinkmann metric (\ref{b1}) requires $\bomega$ to be constant.
This follows from the symmetry of the profile function, since 
\begin{equation}
\dot{\boldsymbol \omega}_{ab} = E^T \left(\dot{\omega} + \Omega \omega + \omega \Omega^T\right)E = 0 \ ,
\label{b199}
\end{equation}
using (\ref{b122}).
We emphasise however that with this proviso, (\ref{b18}) still describes the same spacetime
as the Brinkmann metric (\ref{b1}) and the equivalent standard Rosen metric. It is simply expressed in terms
of a different choice of the non-unique Rosen coordinates adapted to the description of twisted null
congruences.  

The Rosen metric (\ref{b18}) will be the basis of many further developments later in the paper, especially in the
context of the special class of homogeneous plane waves. First, however, we look in more detail at the
construction of the zweibeins and the origin of twist in the corresponding null congruences.

\subsection{Origin of twist}\label{sect2.3}

In order for (\ref{b3}) to be a solution of the geodesic equations, the zweibein $E^i{}_a(u)$ must satisfy 
the oscillator equation
\begin{equation}
\ddot{E}^i{}_a - h^i{}_j E^j{}_a = 0 \ .
\label{b19}
\end{equation}
Viewed as a second-order differential equation for vectors labelled by the index $i$, there are four
linearly independent solutions (therefore a total of 8 solutions for the components) which we split
into two sets $f^i_{(r)}(u)$ and $g^i_{(r)}(u)$, $r = 1,2$, satisfying the canonical boundary conditions\footnote{
It is interesting to note the correspondence with the quantities ${\bf A}(x,x')$ and ${\bf B}(x,x')$ 
introduced, for example, in (3.18) of ref.\cite{Hollowood:2009qz} to characterise geodesic deviation. Here, 
${\bf A}$ and ${\bf B}$ satisfy `parallel' and `spray' boundary conditions in correspondence 
with $f$ and $g$. Note also that in this paper we have chosen the opposite sign convention
$h_{ij} = - R_{uiuj}$ from \cite{Hollowood:2009qz} for the Brinkmann profile function. 
Also note that our $f^i_{(r)}(u)$, $g^i_{(r)}(u)$ are the functions denoted
$b_i^{(k)}(x_0^+)$, $b_i^{*(k)}(x_0^+)$ in \cite{Blau:2002js}.}
\begin{align}
f^i_{(r)}(0) &= \d^i{}_r \ , ~~~~~~~~~~ \dot{f}^i_{(r)}(0) = 0 \ , \nonumber \\
g^i_{(r)}(0) &= 0 \ , ~~~~~~~~~~~~ \dot{g}^i_{(r)}(0) = \d^1{}_r \ .
\label{b20}
\end{align}
An important role in what follows is played by the Wronskian. For example, choosing two solutions
$f^i_{(r)}(u)$ and $g^i_{(s)}(u)$, their Wronskian is\footnote{
The Wronskian $W_{rs}$ is independent of $u$ by virtue of the fact that $f^i_{(r)}$ and $g^i_{(s)}$ satisfy 
the oscillator equation and that $h_{ij}$ is symmetric, since
\begin{equation*}
\dot{W} = f^i_{(r)} h_{ij} g^j_{(s)} - h_{ij} f^j_{(r)} g^i_{(s)} = 0 \ .
\end{equation*}
$W_{rs}$ may therefore be evaluated at any value of $u$, and the result (\ref{b21}) follows 
immediately by using the boundary conditions (\ref{b20}) at $u=0$.}
\begin{equation}
W_{rs} = \sum_i \left(f^i_{(r)} \dot{g}^i_{(s)}  - \dot{f}^i_{(r)} g^i_{(s)} \right) = \d_{rs} \ ,
\label{b21}
\end{equation}
whereas the Wronskian of two $f^i_{(r)}$ or two $g^i_{(r)}$ vanishes.

The zweibein $E^i{}_a$ which determines the geodesics is an arbitrary linear combination of the
$f^i_{(r)}$ and $g^i_{(r)}$, {\it i.e.}~the zweibeins use {\it half} of the complete set of solutions
to the oscillator equation. It follows that the nature of the null congruence depends on the particular
linear combination chosen.  This brings us to the key point. The Wronskian associated with
a particular choice $E^i{}_a(u)$ for the zweibein is
\begin{align}
W_{ab} &= \left(E^T \dot{E}\right)_{ab} - \left(\dot{E}^T E\right)_{ab} \nonumber \\
&= \left(E^T \left(\Omega - \Omega^T\right) E\right)_{ab} \nonumber \\
&= \left({\boldsymbol\Omega} - {\boldsymbol\Omega}^T\right)_{ab} \nonumber \\
&= 2\, {\boldsymbol\omega}_{ab} \ .
\label{b22}
\end{align}
in other words, the twist ${\boldsymbol\omega}_{ab} = \left(E^T \omega E\right)_{ab}$ 
in the Rosen metric.

This identification of the twist with the Wronskian of the zweibeins explains how the formalism
developed here generalises the corresponding discussion in \cite{Blau:2002js}, especially in the
appendix where the transformation between Brinkmann and Rosen coordinates
and the link with Killing vectors was explored in detail. In particular, the `somewhat
mysterious' symmetry condition $\dot{E}E^T = E \dot{E}^T$, already associated in \cite{Blau:2002js} 
with the vanishing of the Wronskian  (corresponding to choosing the zweibeins using oscillator
solutions generating a maximal set of commuting Killing vectors)
is now seen to be the restriction to congruences with vanishing twist. Indeed, the symmetry
condition {\it must} hold for consistency in passing from the usual form of the Rosen metric 
to Brinkmann since the twist term involving ${\boldsymbol\omega}_{ab}$ in the generalised
Rosen metric is omitted, thereby assuming vanishing twist {\it a priori}.

As anticipated above, the oscillator equation solutions $f^i_{(r)}$ and $g^i_{(r)}$ also play
a key role in constructing the Killing vectors characterising the extended Heisenberg
isometry algebra of the special class of homogeneous plane waves,
which we discuss in section \ref{sect 3}.

\subsection{Newman-Penrose tetrad and the Penrose limit}\label{sect2.4}

The standard Newman-Penrose basis for a plane wave spacetime is the null tetrad 
$\ell^\m$, $n^\m$, $m^\m$, $\bar{m}^\m$ satisfying $\ell . n = -1$,
$m.\bar{m}=1$, 
built around $\ell_\m = \partial_\m u$, {\it i.e.} the normal to the null hypersurfaces $u = {\rm const.}$
In terms of this basis, the metric may be written
\begin{equation}
g_{\m\n} = -\ell_\m n_\n - \ell_\n n_\m + m_\m \bar{m}_\n + m_\n \bar{m}_\m \ .
\label{b23}
\end{equation}
A straightforward construction then gives,
\begin{equation}
\ell^\m = \begin{pmatrix}0\\1\\\bf{0}\end{pmatrix} \ , \hskip1.5cm
n^\m = \begin{pmatrix}-1\\\tfrac{1}{2} x^i h_{ij} x^j\\\bf{0}\end{pmatrix} \ , \hskip1.5cm
m ^\m = \frac{1}{\sqrt2}\begin{pmatrix}0\\0\\\d^{i1} + i\d^{i2}\end{pmatrix} \ .
\label{b24}
\end{equation}

The only non-vanishing NP curvature scalars (see, {\it e.g.}~\cite{Chandrasekhar:1985kt}) are
\begin{equation}
\Phi_{22} = -\tfrac{1}{2} R_{\m\n} n^\m n^\n = - \tfrac{1}{2} R_{uu} = \tfrac{1}{2} \tr\,h_{ij} \ ,
\label{b25}
\end{equation}
and 
\begin{equation}
\Psi_4 = -C_{\m\n\r\s}n^\m\bar{m}^\n n^\r \bar{m}^\s = - C_{uiuj} \bar{m}^i \bar{m}^j 
= \tfrac{1}{2}(h_{11} - h_{22}) - i h_{12} \ , 
\label{b26}
\end{equation}
where notably both $\Phi_{00}$ and $\Psi_0$ vanish.  This characterises a Petrov type N spacetime.

In previous work \cite{Hollowood:2009qz},
it proved useful to construct a Newman-Penrose basis adapted to the 
geodesic $\gamma$. Here, taking $k^\m$ from (\ref{b6}), we may define the basis vectors as\footnote{
For reference,
\begin{align*}
L_\m &= k_\m = \begin{pmatrix} 
\tfrac{1}{2}x\left(h - \Omega^T \Omega\right)x, &1, &\left(\Omega x\right)_i \\
\end{pmatrix} \ , ~~~~
N_\m = \begin{pmatrix} -1, &0, &{\bf 0} \\ \end{pmatrix} \ , \nonumber \\
M_\m &= \tfrac{1}{\sqrt2}\begin{pmatrix} - \left(\Omega x\right)_1 - i \left(\Omega x\right)_2, & 0, 
& \d_{i1} + i \d_{i2} \\\end{pmatrix} \ .
\end{align*}}
\begin{equation}
L^\m = k^\m = \begin{pmatrix} 1 \\
-\tfrac{1}{2} x \left(h + \Omega^T \Omega\right) x \\
(\Omega x)^i\\ \end{pmatrix}\ , \hskip0.4cm
N^\m = \begin{pmatrix} 0 \\  -1 \\ {\bf 0} \end{pmatrix} \ , \hskip0.4cm
M^\m = \frac{1}{\sqrt2}\begin{pmatrix} 0 \\ -\left(\Omega x\right)^1 - i \left(\Omega x \right)^2 \\ 
\d^{i1} + i \d^{i2}\\\end{pmatrix} 
\label{b27}
\end{equation}
and check explicitly that they satisfy the metric conditions $L.N = -1$, ~$M.\bar{M} = 1$.
Importantly, we also impose that they are parallel-transported along the geodesic, {\it i.e.}
\begin{equation}
L^\m D_\m L^\n = 0\ , ~~~~~~~ L^\m D_\m N^\n = 0 \ , ~~~~~~~ L^\m D_\m M^\n = 0 \ .
\label{b28}
\end{equation}
The first merits further comment. From its definition in (\ref{b9}),
it follows that
\begin{equation}
L^\m D_\m L_\n =  \left(\Omega - \Omega^T\right)_{\n\m} L^\m + \tfrac{1}{2} D_\n L^2 \ ,
\label{b28a}
\end{equation}
where the final term vanishes since $L^\m$ is null. Normally this would immediately imply
the vanishing of $L^\m D_\m L_\n$, which (in affine parametrisation) is the geodesic equation for $\c$. 
However, if we allow congruences with twist, $\Omega_{\m\n}$ is not symmetric and this 
is no longer obvious.
Nevertheless, using the results quoted in footnote \ref{fn1} we can check explicitly that
\begin{equation}
\left(\Omega - \Omega^T\right)_{\n\m}L^\m =
\begin{pmatrix} 
0&0&-(\dot{x}\omega)_j \\
0&0&0 \\
-(\omega \dot{x})_i &0 &\omega_{ij}
\end{pmatrix}\,\,
\begin{pmatrix}
1 \\ \dot{v} \\ \dot{x}_j
\end{pmatrix}
~=~0 \ ,
\label{b28b}
\end{equation}
confirming the direct calculation of $L^\m D_\m L_\n$ using the covariant derivative.

Although much of this work 
centred on exploiting plane wave geometries as the Penrose limits 
associated with wave propagation along null geodesics in a more general spacetime,
it is clearly of interest to consider the plane wave in its own right and ask what is its Penrose limit
given a particular geodesic $\c$. In particular, we will allow for $\c$ to be a null geodesic in a 
twisted congruence.

In \cite{Hollowood:2009qz} we introduced an elegant method of determining Penrose limits based
on the Newman-Penrose tetrad formalism. The method relies on the correspondence
of the NP tetrad associated with the chosen geodesic $\c$, with the basis vectors parallel-transported
along $\c$, and Fermi null coordinates (FNCs). The construction in terms of FNCs \cite{Blau:2006ar}
gives perhaps the best insight into the nature and properties of the Penrose limit, 
making clear how it captures the geometry of geodesic deviation.
 In particular, as shown in \cite{Blau:2006ar}, the Brinkmann coordinates describing the Penrose limit 
plane wave are identified as FNCs. The upshot is that the profile function $\hat{h}_{ij}$ of the
Penrose limit metric is given in terms of the components of the Weyl and Ricci tensors
in the NP basis associated with $\c$ as:
\begin{equation}
\hat{h}_{ij} = - \begin{pmatrix}
\tfrac{1}{2} \left(C_{LMLM} + C_{L\bar M L \bar M}\right) + \tfrac{1}{2} R_{LL}  
&-\tfrac{i}{2}\left( C_{LMLM} - C_{L\bar M L \bar M}\right) \\
{}&{} \\
-\tfrac{i}{2}\left( C_{LMLM} - C_{L\bar M L \bar M}\right) 
&-\tfrac{1}{2} \left(C_{LMLM} + C_{L\bar M L \bar M}\right) + \tfrac{1}{2} R_{LL} 
\end{pmatrix} \ .
\label{b29}
\end{equation}

The next step is to write the NP basis $L^\m, N^\m, M^\m, \bar{M}^\m$ in terms of the standard basis 
introduced above. We find,
\begin{align}
L^\m &= - \tfrac{1}{2} x \Omega^T \Omega x\, \ell^\m - n^\m +
\left[ \tfrac{1}{\sqrt2}\left(\left(\Omega x\right)_1 - i \left(\Omega x\right)_2 \right) m^\m + {\rm h.c.}\right] \ ,
\nonumber\\
N^\m &= - \ell^\m \ , \nonumber \\
M^\m &= - \tfrac{1}{\sqrt2}\left(\left(\Omega x\right)_1 + i \left(\Omega x\right)_2 \right) \ell^\m + m^\m \ .
\label{b30}
\end{align}
Then, recalling that only the curvatures $\Phi_{22} = -\tfrac{1}{2} R_{nn}$ and $\Psi_4 = - C_{n\bar m n \bar m}$
are non-vanishing in this type N spacetime, we can evaluate (\ref{b29}). We find the elegant general result
\begin{equation}
\hat{h}_{ij} = \begin{pmatrix}
 {\rm Re}\,\Psi_4 + \Phi_{22}  & - {\rm Im}\,\Psi_4 \\
{}&{} \\
-  {\rm Im}\,\Psi_4 & -{\rm Re}\,\Psi_4 + \Phi_{22}
\end{pmatrix} \ ,
\label{b31}
\end{equation}
that is,
\begin{equation}
\hat{h}_{ij} =  h_{ij} \ .
\label{b32}
\end{equation}
This shows the satisfying result that the Penrose limit metric is the same as for the original plane wave.

An important point is that this construction of the Penrose limit does not appear to involve the twist
directly. This is to be expected. The Penrose limit encodes the geometry of geodesic deviation around
a chosen geodesic $\c$ and so depends on the background spacetime and the geodesic $\c$.
On the other hand, twist is a property of the congruence, not an individual geodesic. So while the Penrose 
limit may depend on $\c$, twist itself should play no r\^ole.

\section{Homogeneous Plane Waves and Isometries I -- Brinkmann}\label{sect 3}

As described in section \ref{sect 1}, homogeneous plane waves fall into two classes \cite{Blau:2002js}
specified by particular forms for the profile function $h_{ij}(u)$. Here, we focus on the first,
for which the metric is
\begin{equation}
ds^2 = 2 du dv + \left(e^{\e u}\, h_0 \,e^{-\e u}\right)_{ij} x^i x^j du^2 + \left(dx^i\right)^2 \ ,
\label{c1}
\end{equation}
where $\e_{ij}$ is the usual antisymmetric symbol, $\e_{ij} = \begin{pmatrix} 0&1\\-1&0\end{pmatrix}$,
and $h_0$ is a constant symmetric $2 \times 2$ matrix, which we may take to be diagonal.
Notice immediately that with this form,
\begin{equation}
\dot{h}(u) = \left[\e,h(u)\right] \ .
\label{c2}
\end{equation}
The special case where ${\rm tr}\,h_0 = 0$ is Ricci-flat and is known in the literature as
Ozsv{\'a}th-Sch{\"u}cking spacetime. It is ``anti-Machian''  in the sense that it 
is a geodesically complete, singularity-free, vacuum solution of Einstein's equations which nevertheless
is not Minkowski spacetime \cite{OS}.

Compared with the symmetric space plane wave, or Cahen-Wallach space, for which $h_{ij} = {\rm const.}$,
the metric (\ref{c1}) has an extra isometry. We therefore begin with a discussion of the Killing vectors
and isometry algebra for this metric. 

\subsection{Killing vectors and the isometry algebra}\label{sect3.1}

The isometries of a Riemannian manifold are generated by
Killing vector fields $K$, which are defined such that the Lie derivative of the metric $g_{\m\n}$ with respect 
to $K$ vanishes, {\it i.e.}
\begin{align}
{\mathcal L}_K g_{\m\n} &\equiv K^\r g_{\m\n,\r} + K^\r{}_{,\m} g_{\r\n} + K^\r{}_{,\n} g_{\m\r} \nonumber \\
&= D_\m K_\n + D_\n K_\m \nonumber \\
&= 0 \ .
\label{c3}
\end{align}
It follows that the quantity $K^\m dx_\m/d\l$ is conserved along a geodesic $x^\m(\l)$, where
$\l$ is the affine parameter, {\it i.e.}
\begin{equation}
\frac{d}{d\l} \left(K^\m \frac{dx_\m}{d\l}\right) = 0 \ .
\label{c4}
\end{equation}

The commutators of the Killing vectors define the isometry algebra of the spacetime.
In the case of a general plane wave, with arbitrary profile function $h(u)$, this is the 
Heisenberg algebra for generators $Q_r, P_r$ and $Z$:
\begin{align}
\left[Q_r,Q_s\right] &= 0 \ ,~~~~~~~~\left[P_r,P_s\right] = 0 \ ,~~~~~~~~
\left[Q_r, P_s\right] = \d_{rs} Z \ , \nonumber \\
\left[Z,Q_r\right] &= 0 \ , ~~~~~~~~\left[Z,P_r\right] = 0 \ ,
\label{c5}
\end{align}
while for the homogeneous plane wave (\ref{c1}) this is extended with a further generator $X$, 
related to $u$-translations, satisfying
\begin{align}
\left[X, Q_r\right] &=  \e_{rs} Q_s  +  h_{rs}(0) P_s \ , \nonumber \\
\left[X, P_r\right] &=  Q_r  +  \e_{rs} P_s \ , \nonumber \\
\left[X, Z\right] &= 0 \ .
\label{c6}
\end{align}
Clearly, omitting the $\e_{rs}$ terms recovers the isometry algebra for the Cahen-Wallach spacetime
with $h_{rs}(0)$ identified as the constant profile function $h_{ij}$. 

To see how this arises, consider the coordinate transformations which leave the Brinkmann plane
wave metric invariant. First, the metric is evidently invariant under translations in $v$, with corresponding 
Killing vector $K_Z = \partial_v$, {\it i.e.}
\begin{equation}
v \rta v + \a \ , ~~~~~~~~ K_Z = \partial_v \ , ~~~~~~~~ 
K^\m_Z = \begin{pmatrix}0\\1\\ {\bf 0}\end{pmatrix} \ .
\label{c7}
\end{equation}
We can also check that for arbitrary $h_{ij}(u)$, there is an invariance under
\begin{equation}
u \rta u \ , ~~~~~~~~ v \rta v - \a \dot{F}^i x_i \ , ~~~~~~~~x^i \rta x^i + \a F^i \ ,
\label{c8}
\end{equation}
provided $F^i$ is a solution of the oscillator equation, $\ddot{F}^i = h^i{}_j F^j$.
A convenient choice is to use the canonical solutions $f^i_{(r)}$, $g^i_{(r)}$ given above to define
the generators $Q_r$, $P_r$ respectively, with corresponding Killing vectors:
\begin{align}
K_Q &= - \dot{f}^i_{(r)} x_i \partial_v + f^i_{(r)} \partial_i \ , ~~~~~~~~
K^\m_Q = \begin{pmatrix} 0\\ -\dot{f}^i_{(r)} x_i \\ f^i_{(r)} \end{pmatrix} \ , \nonumber \\
K_P &= - \dot{g}^i_{(r)} x_i \partial_v + g^i_{(r)} \partial_i \ , ~~~~~~~~
K^\m_Q = \begin{pmatrix} 0\\ -\dot{g}^i_{(r)} x_i \\ g^i_{(r)} \end{pmatrix} \ .
\label{c9}
\end{align}
The corresponding conserved quantities are easily identified. For example, taking $u$ as the affine paramter,
we have
\begin{equation}
K^\m_Q \dot{x}_\m \equiv g_{\m\n} K^\m_Q \dot{x}^\n = -\dot{f}^i_{(r)} x_i + f^i_{(r)} \dot{x}_i \ ,
\label{c10}
\end{equation}
and clearly, 
\begin{equation}
\frac{d}{d u}\left(K^\m_Q \dot{x}_\m \right) = - \ddot{f}^i_{(r)} x_i + f^i_{(r)} \ddot{x}_i = 0 \ ,
\label{c11}
\end{equation}
using the oscillator equation for $f^i_{(r)}$ and the geodesic equation (\ref{b2}) for $x^i(u)$.

Finally, for the homogeneous plane wave (\ref{c1}), there is a further invariance involving translations
in $u$, {\it viz.}
\begin{equation}
u \rta u + \a \ , ~~~~~~~~  v \rta v \ , ~~~~~~~~  x^i \rta x^i + \a\, \e^i{}_j x^j \ , 
\label{c12}
\end{equation}
which is easily checked using the relation (\ref{c2}) for $\dot{h}_{ij}(u)$. The corresponding Killing vector
is 
\begin{equation}
K_X = \partial_u + \left(\e x\right)^i \partial_i \ , ~~~~~~~~ 
K^\m_X = \begin{pmatrix} 1\\ 0 \\ (\e x)^i \end{pmatrix} \ .
\label{c13}
\end{equation}
The conserved quantity here is 
\begin{align}
K^\m_X \dot{x}_\m &= g_{uu} + \dot{v} + \dot{x} \e  x  \nonumber \\
&= \tfrac{1}{2} x \left(h - \Omega^T \Omega - 2 \e \Omega\right) x \ ,
\label{c14}
\end{align}
and it can easily be checked using the formulae in section 2 that this is conserved along the geodesic.
The invariance of the metric under the associated Lie derivative is also readily confirmed, for example:
\begin{align}
{\mathcal L}_{K_X} g_{uu} &= K^u_X g_{uu,u} + K^i_X g_{uu,i} ~=~
x \dot{h} x + (\e x)^i \left( h_{ij} x^j + x^j h_{ji}\right) \nonumber \\
&= x \left( \dot{h} + \left[h,\e\right]\right) x ~=~ 0 \ .
\label{c15}
\end{align}

The commutation relations follow directly from the form of the Killing vectors.
For example, 
\begin{align}
\left[K_{Q_r},K_{P_s}\right] &= f^i_{(r)} \partial_i \left( - \dot{g}^j_{(s)} x_j \right) \partial_v
- g^j_{(s)} \partial_j \left(-\dot{f}^i_{(r)} x_i\right) \partial_v \nonumber \\
&= - \d_{ij}\left( f^i_{(r)} \dot{g}^j_{(s)} - \dot{f}^i_{(r)} g^j_{(s)} \right) \partial_v  \nonumber \\
&= - W_{rs}\left(f,g\right) K_Z \ .
\label{c16}
\end{align}
Since the Wronskian is $u$-independent, it may be evaluated at $u=0$ where we may use the 
canonical boundary conditions (\ref{b20}) for the solutions $f^i_{(r)}$, $g^i_{(s)}$ specifying the
Killing vectors $K_{Q_r}$ and $K_{P_s}$ respectively. This gives $W_{rs}(f,g) = \d_{rs}$, as given in 
(\ref{c5}).

Next, we readily find
\begin{equation}
\left[ K_X, K_{Q_r}\right] = - \dot{\F}^i_{(r)} x_i \partial_v + {\F}^i_{(r)} \partial_i \ ,
\label{c17}
\end{equation}
where 
\begin{equation}
{\F}^i_{(r)} = \dot{f}^i_{(r)} - \e^i{}_j f^j_{(r)}  \ .
\label{c18}
\end{equation}
The r.h.s.~is clearly of the same form as the Killing vectors $K_Q$ and $K_P$ and can be written as a 
linear combination of them \cite{Blau:2002js}. To determine this, note that since ${\F}^I_{(r)}$ satisfies the oscillator 
equation,\footnote{From the definition (\ref{c18}), 
\begin{equation*}
\dot{\F}^i_{(r)} = h^i{}_j f^j_{(r)} -\e^i{}_j \dot{f}^J_{(r)} \ ,
\end{equation*}
and so, using(\ref{c2}),
\begin{align*}
\ddot{\F}^i_{(r)} &= \left[\e,h\right]_{ij} f^j_{(r)} + h^i{}_j \dot{f}^j_{(r)} - \e^i{}_j h^j{}_k f^k_{(r)} \nonumber \\
&=h^i{}_j \left( \dot{f}^j_{(r)} - \e^j{}_k f^k_{(r)}\right) ~=~ h^i{}_j {\F}^j_{(r)} \ .
\end{align*}
}
it can be written as a linear combination of the basis set $f^i_{(r)}$, $g^i_{(r)}$ as
\begin{equation}
{\F}^i_{(r)} = a_{rs} f^i_{(s)} + b_{rs} g^i_{(s)} \ .
\label{c19}
\end{equation}
The constant coefficients $a_{rs}$ and $b_{rs}$ are determined by evaluating ${\F}^i_{(r)}$ and
$\dot{\F}^i_{(r)}$ at $u=0$ and using the boundary conditions for $f^i_{(r)}$ and $g^i_{(r)}$. 
This gives
\begin{equation}
{\F}^i_{(r)} = \e_{rs} f^i_{(s)} + h_{rs}(0) g^i_{(s)} \ ,
\label{c20}
\end{equation}
and so we find
\begin{equation}
\left[K_X, K_{Q_r}\right] = \e_{rs} K_{Q_s} + h_{rs}(0) K_{P_s} \ ,
\label{c21}
\end{equation}
as shown in (\ref{c6}). The corresponding result for $\left[K_X, K_{P_r}\right]$ follows similarly.

Notice for future use that the form of (\ref{c16}) involving the Wronskian of the two solutions
characterising the $Q$ or $P$ type Killing vectors, and the form of the solution ${\F}$ in their 
commutator with $K_X$ in (\ref{c17}), did not depend at that stage on the specific choice of
the functions $f^i_{(r)}$ and $g^i_{(r)}$, but would hold for Killing vectors defined with
any solutions of the oscillator equation. Later, we will consider the commutation relations
of Killing vectors with other choices of these solutions, particularly with the related 
construction based on the Rosen metric (\ref{b18}).

\subsection{Homogeneous plane wave as a coset space}

As the name indicates, homogeneous plane waves are example of homogeneous spaces
and as such can be described as a coset space $G/H$, where $G$ is the isometry group
and $H$ is the isotropy subgroup. In the model considered here, the isometry group 
$G$ is generated by the set $\left\{X, Z, Q_r, P_r\right\}$ describing the 
extended Heisenberg algebra (\ref{c5}), (\ref{c6}). The isotropy group may be taken as 
$H = \left\{P_r\right\}$.  The elements of the coset space $G/H$ are then in one-to-one 
correspondence with the four-dimensional manifold described by the metric (\ref{c1}). 

The metric for the homogeneous plane wave of Ozsv\'ath-Sch\"ucking type (\ref{c1})
may be constructed from a knowledge of the isometry algebra
$\mathfrak{g}$ of  (\ref{c5}), (\ref{c6}) in a 
standard way. (See \cite{Papadopoulos:2002bg, Blau:2003rt} 
for an analysis of the singular homogeneous plane wave
(\ref{sect 1}) and \cite{Boulware:1981ns, Shore:1988mn} for discussions of the general formalism.)
The starting point is to regard the isometry group $G$ as a principal $H$-fibre bundle
over the coset manifold $G/H$ and define a section $\ell(x) \in G$ in terms of the
`broken' generators ({\it i.e.}~those generators in $G$ but not in $H$) as follows:
\begin{equation}
\ell(x) = e^{u X} e^{v Z} e^{y.Q} \ ,
\label{c22}
\end{equation}
where $x^\m = (u,v,y^i)$ are coordinates on $G/H$ and we abbreviate $y.Q = y^i Q_r \d^r{}_i$. 
Notice that the choice of section is not unique -- different choices, for example in the ordering
of the factors in (\ref{c22}), correspond to different coordinate choices. 

We then construct the Maurer-Cartan 1-form $\ell^{-1} d\ell \in
{\mathfrak{g}}$ and expand in terms of the generators of $G$ as
\begin{equation}
\ell^{-1} d\ell = 
{\boldsymbol e}^X X + {\boldsymbol e}^Z Z + {\boldsymbol e}^r Q_r + 
{\boldsymbol\omega}^r P_r \ ,
\label{c23}
\end{equation}
where ${\mathbf e}^A = e^A{}_\m dx^\m$ are the frame 1-forms on $G/H$ 
($e^A{}_\m$ are the corresponding vielbeins) 
and ${\boldsymbol\omega}^i$ is a local $H$-connection. 
The metric for the coset space $G/H$ is then 
\begin{equation}
ds^2 = g_{AB} e^A{}_\m e^B{}_\n dx^\m dx^\n \ , 
\label{c24}
\end{equation}
where $g_{AB}$ is a $G$-invariant metric, which may be chosen to reflect the Minkowski 
light-cone coordinates, $g_{XZ} = g_{ZX} = 1$, $g_{rs} = \d_{rs}$.

Following through this construction for the algebra (\ref{c5}), (\ref{c6}), we first write
\begin{equation}
\ell^{-1} d\ell = e^{-y.Q} X e^{y.Q} du + Z dv + Q.dy \ ,
\label{c25}
\end{equation}
where we have used the commutators $\left[Z,X\right] = 0$ and $\left[Z,Q_r\right]=0$.
Now we need the general result 
\begin{equation}
e^{-y.Q} X e^{y.Q} = X + \left[X,y.Q\right] + \tfrac{1}{2} \left[\left[X,y.Q\right],y.Q\right] + \ldots
\label{c26}
\end{equation}
and find, using the commutators
\begin{align}
\left[X,y.Q\right] &= - y.\e.Q + y.h(0).P  \nonumber \\
\left[\left[X,y.Q\right],y.Q\right] &= y.h(0).y Z \ ,
\label{c27}
\end{align}
that the series then terminates since $\left[Z,Q_r\right]=0$. This leaves
\begin{equation}
\ell^{-1} d\ell = \left(X - y.\e.Q + y.h(0).P - \tfrac{1}{2} y.h(0).y Z\right) du + Zdv + Q.dy \ .
\label{c28}
\end{equation}
We therefore identify the frame and connection 1-forms as
\begin{align}
{\boldsymbol e}^X &= du  \nonumber \\
{\boldsymbol e}^Z &= \tfrac{1}{2}y^i h_{ij}(0) y^j du  + dv \nonumber \\
{\boldsymbol e}^r &= \left(\e^i{}_j y^j du  + dy^i\right) \d^r{}_i \nonumber \\
{\boldsymbol \omega}^r &= \left(h^i{}_j(0) y^j du \right)\d^r{}_i \ .  
\label{c29}
\end{align}

The metric is then given by (\ref{c24}) as
\begin{equation}
ds^2 = 2 du dv + y^i h_{ij}(0) y^j du^2 + 
\left(dy^i + \e^i{}_j y^j du \right) \left(dy_i + \e_{ik}y^k du\right) \ .
\label{c30}
\end{equation}
The final step to recover the Brinkmann metric in standard form is to make the
change of variable $y = e^{-\e u} x$ such that $dy = - \e y du + e^{-\e u} dx$.
The metric (\ref{c30}) then becomes simply
\begin{equation}
ds^2 = 2 du dv + x e^{\e u} h_{ij}(0) e^{-e u} x du^2 + \left( dx\right)^2 \ ,
\label{c31}
\end{equation}
recovering the Oszv\'ath-Sch\"ucking metric in Brinkmann form (\ref{c1}).
This confirms that this spacetime is indeed a homogeneous space $G/H$
defined by the extended Heisenberg algebra.

\section{Twisted Rosen Metric for Plane Waves}\label{sect 5}

We now return to the discussion of twisted null congruences in general plane wave spacetimes
in section \ref{sect 2} and focus here on their description in terms of Rosen coordinates.
Our starting point is therefore the plane wave Rosen metric (\ref{b18}) in the form
adapted to twisted congruences:
\begin{equation}
ds^2 = 2 du dV - 2 X^a {\boldsymbol\omega}_{ab} dX^b du + dX^a {\BC}_{ab}(u) dX^b \ .
\label{e1}
\end{equation} 

First, taking the metric (\ref{e1}) on its own merits, we derive the corresponding null geodesics
and optical scalars. This discussion will apply to any plane wave, not necessarily homogeneous.
As in section \ref{sect 3}, we then specialise to homogeneous plane waves and describe the Rosen form
of the Killing vectors and isometry algebra.

\subsection{Rosen metric and geodesics} \label{sect 5.1}

The null geodesic equations following from the Rosen metric (\ref{e1}) are\footnote{
The non-vanishing Christoffel symbols for the metric (\ref{e1}) are
\begin{align*}
\Gamma^v_{uu} &=  X \bomega \BC^{-1} \dot{\bomega} X \ , ~~~~~~~~
\Gamma^v_{ua} = X\bomega \BC^{-1} \bomega  + \tfrac{1}{2} X \bomega \BC^{-1} \dot{C} \ , ~~~~~~~~
\Gamma^v_{ab} = - \tfrac{1}{2} \dot{\BC} \ , \nonumber \\
\Gamma^a_{uu} &= \BC^{-1} \dot{\bomega} X \ , ~~~~~~~~
\Gamma^a_{ub} = \BC^{-1} \bomega + \tfrac{1}{2} \BC^{-1} \dot{\BC} \ . 
\end{align*}
where for generality we have quoted the results including a $u$ dependence for $\bomega$.
}
\begin{align}
\ddot{X} + \BC^{-1} \left(\dot{\BC} + 2 \bomega\right) \dot{X} + \BC^{-1} \dot{\bomega} X 
&= 0  \ , \nonumber \\
\ddot{V} - \tfrac{1}{2} \dot{X} \dot{\BC} \dot{X} + X \bomega \BC^{-1} \left(\dot{\BC} + 2 \bomega\right) \dot{X}
+ X \bomega \BC^{-1} \dot{\bomega} X &= 0 \ .
\label{e2}
\end{align}
We immediately restrict to the  case $\dot{\bomega}=0$ so that the Rosen and Brinkmann metrics are equivalent.
Then, noting that $\dot{\BC} + 2\bomega = 2 \BOmega$, 
these may be written in compact form as
\begin{align}
\BC\ddot{X} + 2 \BOmega \dot{X} &= 0 \ ,  \label{e3a}\\
\ddot{V} - \dot{X} \BOmega \dot{X} - X \bomega \ddot{X} &= 0 \ .
\label{e3b}
\end{align}

The geodesic equation for the transverse coordinates can be written in the convenient form,
\begin{equation}
\frac{d}{du}\left(\BC \dot{X} + 2 \bomega X\right) = 0.
\label{e4}
\end{equation}
The first integrals of the geodesic equations (\ref{e3a}), (\ref{e3b}) are therefore,
\begin{align}
\dot{X} + 2\BC^{-1} \bomega X &= \BC^{-1}\xi  
\label{e5a} \\
\dot{V} &= \eta - \frac{1}{2} \xi X
\label{e5b}
\end{align}
with integration constants $\eta$ and $\xi_a$, the latter equation following following from
\begin{equation}
\ddot{V} = \left( -\frac{1}{2} \dot{X} \BC + X \bomega\right) \ddot{X}  = -\frac{1}{2} \xi \ddot{X} \ .
\label{e6}
\end{equation}
Substituting back into the metric, we see that for a {\it null} geodesic we require $\eta = 0$.

To solve (\ref{e5a}), we introduce the path-ordered exponential ${\cal P}(u,a)$ defined by 
\begin{equation}
{\cal P}(u,a) = {\cal T}_- \exp \int_a^u dt \, 2 \BC^{-1}(t) \bomega \ ,
\label{ee15}
\end{equation}
with $a$ arbitrary, where ${\cal T}_-$ denotes anti-$u$ ordering.\footnote{
Note that the corresponding $u$ ordered exponential ${\cal P}_+$ defined
with ${\cal T}_+$ would satisfy 
\begin{equation*}
\dot{{\cal P}}_+(u,a) =  2 \BC^{-1}(u) \bomega{\cal P}_+(u,a) \ .
\end{equation*}
We will also use the relation,
\begin{equation*}
{\cal P}(u,a) \equiv {\cal T}_- \exp \int_a^u dt \, 2 \BC^{-1}(t) \bomega
~=~ {\cal T}_+ \exp \left[-\int_u^a dt \, 2 \BC^{-1}(t) \bomega\right] \ .
\end{equation*}
}
 Its derivative has the key property,
\begin{equation}
\dot{{\cal P}}(u,a) = {\cal P}(u,a) \, 2 \BC^{-1}(u) \bomega \ .
\label{ee16}
\end{equation}
This is the required integrating factor for (\ref{e5a}). Writing
\begin{equation}
{\cal P} \dot{X} + \dot{{\cal P}} X = {\cal P}\, \BC^{-1} \xi \ ,
\label{ee17}
\end{equation}
and integrating, we therefore find
\begin{equation}
{\cal P}(u,a)\, X(u) - {\cal P}(u',a)\, X(u') =
\int_{u'}^u du_1 \, {\cal P}(u_1,a)\, \BC^{-1}(u_1) \, \xi \ .
\label{ee18}
\end{equation}

This simplifies if we set $a = u$, since ${\cal P}(u,u) = 1$ and we find
\begin{equation}
X(u) = \int_{u'}^u du_1 \,{\cal P}(u_1,u)\, \BC^{-1}(u_1)\, \xi  ~+~ {\cal P}(u',u) X(u') \ .
\label{ee19}
\end{equation}
Now, recalling its definition in (\ref{e5a}), the constant $\xi$ may be evaluated for any 
value of $u$, so in particular we may write
\begin{equation}
\xi = \BC(u') \dot{X}(u') + 2 \bomega X(u') \ .
\label{ee20}
\end{equation}
Substituting into (\ref{ee19}) we then have
\begin{multline}
X(u) = \int_{u'}^u du_1\,{\cal P}(u_1,u)\,\BC^{-1}(u_1)\,\BC(u') \dot{X}(u')  \\ ~+~
\left[{\cal P}(u',u) + \int_{u'}^u du_1\, {\cal P}(u_1,u) \, 2 \BC^{-1}(u_1) \bomega \,\right] X(u') \ .
\label{ee21}
\end{multline}
Finally, using the relation (\ref{ee16}) to simplify the integral in square brackets,
we find the elegant result,
\begin{equation}
X(u) = \int_{u'}^u du_1 \, {\cal P}(u_1,u) \,\BC^{-1}(u_1)\,\,\BC(u') \dot{X}(u') ~+~ X(u') \ .
\label{ee22}
\end{equation}
We can of course now check by explicit differentiation that (\ref{ee22}) does indeed 
satisfy the geodesic equation. 

This form of the solution is particularly useful since it isolates the dependence on the twist
entirely in the path-ordered exponential. It is also convenient at this point to introduce some simplified
notation, which will also make the Rosen description of the Killing vectors and isometries 
in the following section more transparent.\footnote{The notation here is chosen to be as close as
possible to refs.~\cite{Duval:2017els,Zhang:2017rno,Zhang:2017geq} 
to allow an easy comparison with the corresponding results for zero
twist. The zero-twist analogue of the function $p{\cal H}(u)p$ was denoted by $\psi(u)$  
in \cite{Gibbons:1975jb, Hollowood:2008kq}.  }
We therefore define the key function ${\cal H}^{ab}(u)$ as
\begin{equation}
{\cal H}(u) = \int_{u'}^u du_1\, {\cal P}(u_1,u)\,\BC^{-1}(u_1) \ ,
\label{ee23}
\end{equation}
or equivalently,
\begin{equation}
{\cal H}(u) = \int_{u'}^u du_1\, {\cal T}_+\exp\left[-\int_{u_1}^u dt\,2 \BC^{-1}(t) \bomega \right] \ .
\label{ee24}
\end{equation}
By construction, this satisfies
\begin{equation}
\BC \ddot{\cal H} + 2 \BOmega \dot{\cal H} = 0 \ ,
\label{ee25}
\end{equation}
together with the important identity,
\begin{equation}
\BC \dot{\cal H} + 2 \bomega {\cal H} = {\bf 1} \ .
\label{ee26}
\end{equation}

We also denote the integration constants representing the Rosen position and velocity
at the reference point $u'$ by
\begin{equation}
p_a = \BC_{ab}(u') \dot{X}^b(u') \ , \hskip2cm  a^a = X^a(u') \ .
\label{ee27}
\end{equation}
Note though that for non-vanishing twist, the conserved integral of motion is actually
\begin{equation}
\xi = \BC \dot{X} + 2 \bomega X = p + 2 \bomega a \ .
\label{ee28}
\end{equation}
The geodesic solution (\ref{ee22}) is then written in compact form as
\begin{equation}
X(u) = {\cal H}(u) p + a \ .
\label{ee29}
\end{equation}

Now return to the geodesic equations (\ref{e3b}) and (\ref{e5b}) for $V$. 
Integrating (\ref{e5b}) and fixing the integration constant at $u=u'$, immediately gives
\begin{equation}
V(u) = -\frac{1}{2} \xi \bigl (X(u) - X(u') \bigr) + \eta(u-u') + V(u') \ .
\label{ee30}
\end{equation}
Rewriting in the simplified notation above, and with $d = V(u')$, we may then show
\begin{equation}
V(u) = - \frac{1}{2} p {\cal H}(u) p - p {\cal H }^T(u) \bomega a + \eta (u-u') + d \ .
\label{ee31}
\end{equation}
Compared with the twist-free case \cite{Duval:2017els,Zhang:2017rno,Zhang:2017geq},   
the twist $\bomega$ therefore enters the geodesic solution for $V$ explicitly as well as 
implicitly through the form of ${\cal H}(u)$.

\subsection{Twisted null congruences and optical scalars}\label{sect 5.2}

The simplest null congruence to consider is the original one described in Brinkmann coordinates in
section \ref{sect 2}, where $V$, $X^a$ are constant. The corresponding tangent vector $\tilde{\Bk}^\m$ 
and covariant vector $\tilde{\Bk}_\m = g_{\m\n} \tilde{\Bk}^\n$  are then
\begin{equation}
\tilde{\Bk}^\m = \begin{pmatrix} 1 \\ 0 \\ {\bf 0} \end{pmatrix} \ , \hskip2cm
\tilde{\Bk}_\m = \begin{pmatrix} 0 \\ 1 \\ \bomega X \end{pmatrix} \ .
\label{e8a}
\end{equation}
Defining $\tilde{\BOmega}_{\m\n} \equiv D_\n \tilde{\Bk}_\m$, we find all the components vanish
except for
\begin{align}
\tilde{\BOmega}_{ab} ~&=~ \partial_b \tilde{\Bk}_a - \Gamma^v_{ba} \tilde{\Bk}_v  \nonumber \\
&=~\bomega_{ab} + \frac{1}{2} \dot{\BC}_{ab} ~=~ \BOmega_{ab} \ .
\label{e8b}
\end{align}
For this congruence, therefore, we recover the natural relation for the twist, 
$\tilde{\bomega} = \bomega$. 

It is interesting to compare this with the null congruence defined with the more general
solutions to the Rosen geodesic equations described above.
These solutions (for $\eta = 0$) define a null congruence with tangent vector
\begin{equation}
\hat {\Bk}^\m ~=~  \begin{pmatrix} \dot{u} \\ \dot{V} \\ \dot{X} \end{pmatrix}
~~=~~ \begin{pmatrix} 1 \\ -\frac{1}{2} \xi \BC^{-1} \left(\xi - 2 \bomega X\right) \\
\BC^{-1}\left(\xi - 2 \bomega X\right) \end{pmatrix} \ .
\label{e7}
\end{equation}
and
\begin{equation}
\hat {\Bk}_\m ~=~ \begin{pmatrix} \dot{V} - X \bomega \dot{X} \\ 1 \\ \BC \dot{X} +\bomega X
\end{pmatrix}
~~=~~ \begin{pmatrix} -\frac{1}{2}\left(\xi + 2 X \bomega\right) \BC^{-1} \left(\xi - 2 \bomega X\right) \\
1 \\ \xi - \bomega X \end{pmatrix} \ ,
\label{e8}
\end{equation}
and we readily check $\hat {\Bk}^2 = 0$.

The optical scalars for this congruence, defined entirely within the 
Rosen metric framework, are constructed as above. Specifically, we have\footnote{The complete result 
for $\hat{\BOmega}_{\m\n} \equiv D_\n \hat{\Bk}_\m$ can be written analogously to footnote \ref{fn1}
in section \ref{sect 2} as   
\begin{equation*}
\hat{\BOmega}_{\m\n} ~\equiv~ D_\n \hat {\Bk}_\m  ~=~ 
\begin{pmatrix} \dot{X} \BOmega^T \dot{X} & ~0~  &- \dot{X} \BOmega^T \\
0 & ~0~ & {\bf 0} \\
- \BOmega^T \dot{X} &  ~{\bf 0}~ & \BOmega^T \end{pmatrix} \ .
\end{equation*}
}
\begin{align}
\hat{\BOmega}_{ab} ~&\equiv~ D_b\hat  {\Bk}_a  ~=~ 
\partial_b \hat{\Bk}_a - \Gamma^v_{ba}\hat {\Bk}_v \nonumber \\
& = ~ -\bomega_{ab} + \frac{1}{2} \dot{\BC}_{ab} ~=~ \left(\BOmega^T\right)_{ab} \ .
\label{e9}
\end{align}
With the usual decomposition into optical scalars, $\hat{\BOmega} = \tfrac{1}{2}\hat{\boldsymbol \theta} {\bf 1}
+ \hat{\boldsymbol \sigma} + \hat{\bomega}$, we find that the twist for this congruence is given by
$\hat{\bomega} = - \bomega$. In section \ref{sect 6}, we see how these two different congruences
are related to the $Q$ and $P$ generators of the isometry algebra, and how this different result
for the Rosen twists $\tilde{\bomega}$ and $\hat{\bomega}$ is reflected in the commutation relations.

\subsection{Wave equation and twist}

We now consider solutions of the wave equation in the gravitational plane wave background
described by the twisted Rosen metric (\ref{e1}).  The d'Alembertian in this metric is
\begin{equation}
\Box = \frac{1}{\sqrt{-g}} \partial_\m \left(\sqrt{-g} g^{\m\n} \partial_\n\right) \ ,
\label{e10}
\end{equation}
where
\begin{equation}
g_{\m\n} = \begin{pmatrix} 0 & ~1~ & -X\bomega \\
1 & ~0~ & {\bf 0} \\
\bomega X & ~{\bf 0}~ & \BC \end{pmatrix}  \ , 
\hskip2cm
g^{\m\n} = \begin{pmatrix} 0 & ~1 & {\bf 0} \\
1 & ~ -X\bomega \BC^{-1} \bomega X~ & X\bomega \BC^{-1} \\
{\bf 0} & ~- \BC^{-1}\bomega X~ & \BC^{-1} \end{pmatrix} \ ,
\label{e11}
\end{equation}
and $\sqrt{-g} = \det\, \BC$.  It follows that 
\begin{equation}
\partial_u \log\sqrt{-g} = \frac{1}{2} \tr\, \BC^{-1} \dot{\BC} 
= \frac{1}{2} \tr\left(\Omega + \Omega^T\right) = \theta \ ,
\label{e12}
\end{equation}
where $\theta$ is the expansion scalar (\ref{b10}). Collecting terms, we find the following 
elegant form,
\begin{equation}
\Box ~=~ 2\, \partial_u \,\partial_V +
\big(\partial_X + X \bomega\, \partial_V \big) \BC^{-1}
\big(\partial_X - \bomega X\, \partial_V\big) 
+ \theta\, \partial_V \ .
\label{e13}
\end{equation}

Generalising from the twist-free case, we look for solutions $\phi(x)$ of the form
\begin{equation}
\phi(x) = {\cal A}(u)\, F(u,X)\, e^{iV} \ ,
\label{e14}
\end{equation}
where $V$ is the usual phase and the the amplitude factor ${\cal A}(u)$ satisfies
\begin{equation}
\partial_u \log\,{\cal A} = - \frac{1}{2} \theta \ .
\label{e15}
\end{equation}
This shows how the amplitude reflects the overall expansion of the null congruence,
in the way familiar from conventional geometric optics. We then have
\begin{equation}
\Box\,\phi ~=~ {\cal A}(u)\, e^{iV} \, \Big[2 i \partial_u ~+~
\big(\partial_X + i X \bomega \big) \BC^{-1} \big(\partial_X - i \bomega X \big) \Big] \,F(u,X) \ .
\label{e16}
\end{equation}
Clearly, $\phi(x)$ is a solution of the wave equation if $F(u,X)$ solves
\begin{equation}
\Bigl[2 i\,\partial_u ~+~ \left(\partial_X + i X \bomega\right) \BC^{-1} 
\left(\partial_X - i \bomega X\right) \Big]\,F(u,X) ~=~0 \ ,
\label{e17}
\end{equation}
which we can write as 
\begin{equation}
\Big[ 2i \,\partial_u ~+~ D_a (\BC^{-1})^{ab} D_b \Big] \, F(u,X) ~=~ 0 \ ,
\label{e18}
\end{equation}
with $D_a = \partial_a - i \bomega_{ab}X^b$. These $D_a$ are in essence covariant 
derivatives with abelian connection $\bomega_{ab}X^b$.  Provided that $\dot{\bomega} = 0$,
this corresponds to a constant field strength $F_{ab} =  -2 \bomega_{ab}$. 

We see immediately that this takes the form of a heat equation (for $u$ imaginary) 
and so may readily be solved using techniques from the heat kernel or Schwinger
proper-time formalisms. We can therefore write a standard form of solution as
\begin{equation}
F(u,X) ~=~ \Phi(X,Y) \, \exp\left[\frac{i}{2} (X-Y)^a A_{ab}(u) (X-Y)^b + B(u)\right] \ ,
\label{e19}
\end{equation}
where $\Phi(X,Y)$ is the path-dependent phase factor,
\begin{equation}
\Phi(X,Y) ~=~ \exp\left[ i \int_Y^X dZ^a\, \bomega_{ab}Z^b \right] \ .
\label{e20}
\end{equation}
The r\^ole of the phase factor in this type of solution (see, {\it e.g.} ref.~\cite{Shore:1981mj})
is to convert the covariant derivatives to ordinary derivatives and field strengths as they act 
through $\Phi$ to the exponential term in (\ref{e19}). This follows from the key property 
(for the special case of constant field strength),
\begin{equation}
D_a \Phi(X,Y) ~=~ - i \Phi(X,Y) \,\bomega_{ab}\,(X-Y)^b \ ,
\label{e21}
\end{equation}
where we have taken the path from $Y$ to $X$ at constant $u$.

In the present case with $\dot{\bomega} = 0$, however, this geometrically natural construction 
is not needed in its full generality, since choosing $Y=0$ we can show that the phase factor is trivial,
$\Phi(X,0) = 1$. Our ansatz for the solution to (\ref{e18}) is then simply,\footnote{Note that while
a linear term in $X$ may in principle be included in this ansatz, the wave equation requires it to vanish
provided $B$ is real.}
\begin{equation}
F(u,X) ~=~ \exp\left[\frac{i}{2}X^a A_{ab}(u) X^b + B(u) \right] \ ,
\label{e22}
\end{equation}
where, after equating real and imaginary terms, the $u$-dependent functions
$A_{ab}(u)$ and $B(u)$ satisfy
\begin{align}
\frac{d}{du}\, A +  \left( A +\bomega \right) \BC^{-1}\left(A - \bomega \right) ~&=~0 \nonumber \\
\frac{d}{du}\, B + \frac{1}{2} \tr\, \BC^{-1} A ~&=~0 \ .
\label{e23}
\end{align}
The terms involving $A$ and $B$ therefore modify the phase and amplitude of $\phi(x)$ respectively.

Equation (\ref{e23}) is a non-linear, first order ODE for the symmetric matrix 
$A_{ab}(u)$, which we recognise as a Riccati equation. This motivates converting it into a linear,
second order ODE with an appropriate change of variables. We therefore introduce the matrix
$M^{ab}(u)$ such that 
\begin{equation}
\dot{M} = \BC^{-1} \left(A - \bomega\right) \,M \ .
\label{e24}
\end{equation}
Differentiating, and using $\dot{\BC} = 2(\BOmega - \bomega)$ and $\dot{\bomega} = 0$,  we find
\begin{align}
\ddot{M} &= \left[-2 \BC^{-1} \left(\BOmega - \bomega\right) - \BC^{-1}\left(A + \bomega\right)
+ \BC^{-1} \left(A - \bomega\right) \right]\, \dot{M} \nonumber \\
&= - 2 \BC^{-1} \BOmega\, \dot{M} \ ,
\label{e25}
\end{align}
with the term of $O(A^2)$ vanishing by construction. That is, $M$ is the solution of
\begin{equation}
\ddot{M} + 2 \BC^{-1} \BOmega \, \dot{M} ~=~ 0 \ .
\label{e26}
\end{equation}

We recognise this as the defining equation (\ref{ee25}) for the key function ${\cal H}(u)$ 
given in (\ref{ee23}). Inverting (\ref{e24}), we therefore solve for $A_{ab}$ in the form,
\begin{equation}
A  ~=~ \BC\, \dot {\cal H}\, {\cal H}^{-1} + \bomega ~=~  {\cal H}^{-1} -  \bomega\ ,
\label{e27}
\end{equation}
using (\ref{ee26}).
The final solution for the wave equation is then,
\begin{equation}
\phi(x) = \tilde{\cal A}(u)\, \exp\Big[i \big(V + \tfrac{1}{2} X {\cal H}^{-1}(u) X \big) \Big] \ ,
\label{e28}
\end{equation}
with amplitude
\begin{equation}
\tilde{\cal A}(u) = \exp\left[-\frac{1}{2} \int du\, \tr\Big(\theta + \left({\cal H} \BC\right)^{-1}
\Big)\right] \ .
\label{e29}
\end{equation}

It is interesting to consider this solution in the special case of zero twist, where a geodesic
with tangent vector $k^\m$ can be given in terms of the derivative of the phase of
a solution of the wave equation, {\it i.e.} $k_\m = \partial_\m \Theta$.
Of course this is not possible for a congruence with non-vanishing twist, which cannot
be described as a gradient flow. Here, taking $\Theta$ as the phase in (\ref{e28}),
this would give
\begin{equation}
k_\m = \begin{pmatrix}\tfrac{1}{2} X \dot{\cal H}^{-1} X \\ 1 \\  {\cal H}^{-1} X \end{pmatrix}
~=~ \begin{pmatrix} -\tfrac{1}{2} p \BC^{-1} p \\ 1 \\ p\end{pmatrix} \ ,
\label{e30}
\end{equation}
since with $\bomega = 0$, we have $\dot {\cal H} = \BC^{-1}$ and we have set $X = {\cal H}(u) p$.
This reproduces the zero-twist limit of the expression found in (\ref{e8}) directly from the
geodesic equations.

\section{Homogeneous Plane Waves and Isometries II -- Rosen}\label{sect 6}

In this section, we reconsider the symmetries of gravitational plane
waves, and the extended symmetries of homogeneous plane waves, this time from the
point of view of the twisted Rosen metric (\ref{b18}). We derive the explicit
Rosen forms of the isometries and Killing vectors and show how the extended Heisenberg algebra
of section \ref{sect 3} arises entirely within the framework of Rosen coordinates. 

\subsection{Killing vectors}

The first isometry of the Rosen metric (\ref{b18}), which we display here again for ease of reference,
\begin{equation}
ds^2 = 2 du dV - 2 X^a {\boldsymbol\omega}_{ab} dX^b du + dX^a {\BC}_{ab}(u) dX^b \ ,
\label{g1}
\end{equation}
is of course invariance under translations of $V$, with generator $Z$ and Killing vector $K_Z$,
{\it i.e.}
\begin{equation}
u \rta u \ , ~~~~~~~~  V \rta V + \alpha \ , ~~~~~~~~ X \rta X \ ; ~~~~~~~~~~~~ K_Z = \partial_V \ .
\label{g2}
\end{equation}

The basic Heisenberg algebra is completed by the generators $Q_r$, $P_r$ which involve a choice of
independent solutions $F^i{}_{(r)}$, $G^i{}_{(r)}$ respectively of the oscillator equation.  
Unlike section \ref{sect 3}, however, we will not immediately commit to the choice $f,g$ satisfying
the canonical boundary conditions (\ref{b20}). Focusing now on $Q_r$, and converting from the
Brinkmann transformation (\ref{c8}), we expect the Rosen metric to be invariant under the
transformations
\begin{equation}
u \rta u \ , ~~~~~~~~
V \rta V - \a \left[\dot{F}^T E  - F^T E^{T-1}\left(\BOmega - \bomega\right)
\right]X  \ ,  
~~~~~~~~ X \rta X + \a\, F^T E^{T-1} \ .
\label{g3}
\end{equation}
Here,\footnote{
The conversion from Brinkmann to Rosen coordinates,
\begin{equation*}
u = u \ ,  ~~~~~~~~ v = V - \frac{1}{2} X \BOmega X \ ,  ~~~~~~~~ x = E X \ ,
\end{equation*}
suppressing indices, implies
\begin{equation*}
\partial_u = \partial_u \ ,  ~~~~~~~~ \partial_v = \partial_V \ , ~~~~~~~~ 
\partial_x = E^{T-1} \left(\partial_X + \left(\BOmega - \bomega\right)X\,\partial_V\right)  \ .
\end{equation*}
} 
we use an abbreviated matrix notation $F = F^i{}_{(r)}$ and parameters $\a = \a^r$, 
so that written in full the $X$ transformation, for example, is 
$X^a \rta \a^r F_{(r)}{}^i (E^{-1})_i{}^a$.
The corresponding Killing vector is therefore
\begin{equation}
K_{Q_r} =F^TE^{T-1}\,\partial_X - 
\left[\dot{F}^T E  - F^T E^{T-1} \left(\BOmega - \bomega\right) \right]X\,\partial_V \ ,
\label{g4}
\end{equation}
with the analogous form for $K_{P_r}$, {\it viz.}
\begin{equation}
K_{P_r} =G^TE^{T-1}\,\partial_X - 
\left[\dot{G}^T E  - G^T E^{T-1} \left(\BOmega - \bomega\right) \right]X\,\partial_V \ ,
\label{g4a}
\end{equation}

The direct proof that (\ref{g3}) leaves the Rosen metric invariant requires some calculation.
Useful intermediate results include
\begin{equation}
dX \rta dX + \a \left( \dot{F}^T E - F^T E^{T-1} \BOmega^T \right) \BC^{-1}\,du \ ,
\label{g5}
\end{equation}
and 
\begin{align}
dV \rta dV &- \a \left[\dot{F}^T E  - F^T E^{T-1} \left(\BOmega - \bomega\right) \right] \,dX 
\nonumber \\
&- \a \left[\left(\dot{F}^T E  - F^T E^{T-1} \BOmega^T\right)  \BC^{-1} \bomega \right. \nonumber \\
& ~~~~~~~~\left. + \ddot{F}^T E - F^T E^{T-1} \dot{\BOmega} + F^T E^{T-1}  \BOmega^T \BC^{-1} \BOmega
\right] X \, du \ ,
\label{g6}
\end{align}
provided $\dot{\bomega} = 0$, as required for the equivalence of the Brinkmann and Rosen metrics.
Then using the fact that $F$ is an oscillator solution to write $\ddot{F}^T E = F^T E^{T-1} \Bh$, 
together with the analogue of (\ref{b11}),
\begin{equation}
\Bh = \dot{\BOmega} - \BOmega^T \BC^{-1} \BOmega \ ,
\label{g7}
\end{equation}
we can verify that the variation of the Rosen metric indeed vanishes under (\ref{g3}). 

This general form of the $Q_r$ and $P_r$ isometries is not at all evident from inspection
of the Rosen metric. However, we can make particular choices of the oscillator solutions
$F^i_{(r)}$ and $G^i_{(r)}$ which correspond more directly to the natural invariances of the 
metric.

First, we define the generator $Q_a$ by taking $F^i{}_{(r)} = E^i{}_a$,
since by construction the zweibein $E^i{}_a$ is itself a solution of the oscillator equation.
The corresponding isometry, with parameter $c^a$, is then simply
\begin{equation}
u \rta u \ , ~~~~~~~~ V \rta V + c\, \bomega X \ , ~~~~~~~~ X \rightarrow X + c \ ,
\label{g8}
\end{equation}
since in this case $\dot{F}^T E = \BOmega^T$, with Killing vector
\begin{equation}
K_{Q} = \partial_X + \bomega X\,\partial_V \ .
\label{g9}
\end{equation}
The transformation (\ref{g8}) is manifestly a symmetry of the twisted Rosen metric. 
Notice the key point that a simple $X$ translation is not an isometry, but must be 
accompanied by a twist-dependent transformation of $V$.

Next, note that a further oscillator solution is given by $G^i_{(r)} = E^i{}_a H^{ab}$
provided $H^{ab}(u)$ satisfies the equation,
\begin{equation}
\ddot{H} + 2 \BC^{-1} \BOmega \dot{H} = 0 \ .
\label{gg10}
\end{equation}
This follows immediately from
\begin{align}
\ddot{G} &= \ddot{E} H + 2 \dot{E} \dot{H} + E \ddot{H} \nonumber \\ 
&= h G + E \big(\ddot{H} + 2 \BC^{-1} \BOmega H \big) \ .
\label{gg11}
\end{align}
We then look for a corresonding isometry of the form,
\begin{equation}
u \rightarrow u \ , ~~~~~~~~
V \rightarrow V - X N b - \frac{1}{2} b M b \ , ~~~~~~~~
X \rightarrow X + Hb \ ,
\label{gg12}
\end{equation}
with parameter $b_a$, for functions $H^{ab}(u)$, $N_a{}^b(u)$ and $M^{ab}(u)$ to be determined.
Here, we have included a term of $O(b^2)$ to identify a finite transformation, although
this will not be present in the infinitesimal transformations defining the Killing vectors.
To check that this is indeed an invariance of the metric, we substitute (\ref{gg12})
together with 
\begin{equation}
dX \rta dX + \dot{H}b\, du \ ,
\label{gg13}
\end{equation}
and
\begin{equation}
dV \rta dV - dX N b - X \dot{N} b \,du - \frac{1}{2} b \dot{M} b \,du \ ,
\label{gg14}
\end{equation}
into the Rosen metric. A short calculation shows that the metric is invariant provided
$H(u)$ satisfies (\ref{gg10}) and
\begin{align}
N &= \BC \dot{H} + \bomega H \ , \label{gg15} \\
\dot{M} &= \dot{H}^T \big( \BC \dot{H} + 2 \bomega H \big) \ .
\label{gg16}
\end{align}

Notice now that the equation (\ref{gg10}) for $H^{ab}(u)$ is precisely the same as 
that satisfied by the function ${\cal H}^{ab}(u)$ involving the path-ordered integrating factor
introduced in (\ref{ee23}) in the discussion of Rosen geodesics. 
We therefore choose $H \equiv {\cal H}$ and, using the further identity (\ref{ee26}),
define the $P^a$ isometry to be,
\begin{equation}
u \rta u \ , ~~~~~~~~
V \rta V - X \big( {\bf 1} - \bomega {\cal H} \big) b - \frac{1}{2} b \dot{\cal H} b \ , ~~~~~~~
X \rta X + {\cal H}b \ ,
\label{gg17}
\end{equation}
with corresponding Killing vector,
\begin{equation}
K_P = {\cal H}^T\, \partial_X - \left({\bf 1} + {\cal H}^T \bomega\right) X\,\partial_V \ .
\label{gg18}
\end{equation}
It is now easy to check that this precisely reproduces the general form (\ref{g4a}) for the
particular choice $G = E {\cal H}$.

The remaining generator $X$ of the extended Heisenberg algebra only corresponds to an isometry when 
the Rosen metric describes a homogeneous plane wave. This is encoded through the functions
$\BC$ and $\bomega$ in the metric, and therefore implicitly through the zweibein $E$ which is 
determined by the Brinkmann profile function $h(u)$. Proving that this is indeed an isometry therefore
relies on the use of the homogeneous plane wave condition (\ref{c2}), $\dot{h}(u) = \left[\e,h(u)\right]$.

Again starting from the equivalent Brinkmann transformation (\ref{c12}), we expect the Rosen metric to 
be invariant under 
\begin{equation}
u \rta u + \a \ , ~~~~~~~~ V \rta V + \a\,\frac{1}{2} X \BA X \ , ~~~~~~~~ X \rta X + \a\, \BB X \ , 
\label{g10}
\end{equation}
with corresponding Killing vector,
\begin{equation}
K_X = \partial_u + X \BB^T\,\partial_X + \frac{1}{2} X \BA X\,\partial_V \ ,
\label{g11}
\end{equation}
where
\begin{align}
\BB &= -\BC^{-1} \left(\BOmega - \bepsilon\right) \ , \label{g12a}\\
\BA &= \Bh - \BOmega^2 + \left[\BOmega,\bepsilon\right] \ . \label{g12b}
\end{align}
Here, $\BB = \BB^a{}_b$, $\,\BA = \BA_{ab}$ and for clarity we have omitted the $\BC^{-1}$ factors in 
contracting Rosen indices in the expression for $\BA$. 
Notice that while the $V$ transformation only involves its symmetric part,
$\BA$ itself is not symmetric when the twist is non-vanishing.
Instead, we have the useful identity,
\begin{equation}
\frac{1}{2}\left(\BA - \BA^T\right) = \BB^T \bomega + \bomega \BB \ .
\label{g13}
\end{equation}

To check the invariance of the twisted Rosen metric under (\ref{g10}), we need the
$u$-derivatives,
\begin{equation}
\dot{\BB} = - \BC^{-1} \BA \ , ~~~~~~~~~~~~\dot{\BA} = -2\, \bomega \BC^{-1} \BA \ . 
\label{g14}
\end{equation}
Then
\begin{align}
dX &\rta dX + \a\,\BB\,dX - \a\,\BC^{-1} \BA X \,du \ , \nonumber \\
dV &\rta dV + \frac{1}{2}\a\,X\left(\BA + \BA^T\right)\,dX - \a\,X \bomega \BC^{-1} A X\, du \ ,
\label{g15}
\end{align}
and invariance of the metric follows directly.  Again, we see that in contrast to the Brinkmann
description, the enhanced isometry for the homogeneous plane wave involves a compensating
transformation of $V$ as well as $X$ to balance the $u$-translation.

\subsection{Isometry algebra}

Having identified the isometries of the twisted Rosen metric, the next step is to find the
corresponding algebra by evaluating the commutators of the Killing vectors
$K_Z,\,K_Q\, K_P$ and $K_X$.
We perform these calculations initially with arbitrary oscillator solutions defining 
$Q_r\, P_r$, and work entirely in Rosen coordinates.

First, the Killing vector $K_Z = \partial_V$ naturally commutes with all the others,
so we have simply,
\begin{equation}
\left[Z,Q_r\right] = 0 \ , ~~~~~~~~ \left[Z,P_r\right] = 0 \ , ~~~~~~~~ \left[Z,X\right] = 0 \ .
\label{g16}
\end{equation}

Next, we show that
\begin{equation}
\left[Q_r, Q_s\right] = -W_{rs}(F,F) \,Z \ , ~~~~~~~~\left[P_r,P_s\right] = - W_{rs}(G,G) \,Z  \ ,
\label{g17}
\end{equation}
and
\begin{equation}
\left[Q_r, P_s\right] = - W_{rs}(F,G)\, Z \ ,
\label{g18}
\end{equation}
in terms of the appropriate Wronskian, {\it e.g.}
\begin{equation}
W_{rs}(F,G) \,=\, F^T \dot{G} \,-\, \dot{F}^T G \ .
\label{g19}
\end{equation}
This follows readily from the expressions (\ref{g4}), (\ref{g4a}) for the Killing vectors,
which imply
\begin{align}
\left[K_{Q_r}, K_{P_s}\right] &= 
- \left[\left(\dot{G}^T E - G^T E^{T-1}\left(\BOmega - \bomega\right) \right) E^{-1} F \right]_{sr}\, \partial_V
\,-\, \left[~~F \leftrightarrow G~~\right]_{rs}\, \partial_V \nonumber \\
&= - \left(F^T \dot{G} \,-\,\dot{F}^T G\right)_{rs}\, \partial_V \nonumber \\
&= -\, W_{rs}(F,G) \, K_Z \ ,
\label{g20}
\end{align}
with the terms not involving derivatives of $F$ and $G$ cancelling.

The evaluation of the $\left[X,Q_r\right]$ commutator in the Rosen case involves a lengthier
calculation. First, using the definition (\ref{g12a}) for $\BB$, we eventually find
\begin{multline}
\left[K_X, K_{Q_r}\right] ~=~ \left(\dot{F}^T + F^T \epsilon\right) E^{T-1} \, \partial_X \\
+ F^T E^{T-1} \left[ \BOmega^T \BC^{-1} \bomega - 
\left(\BOmega - \bomega\right) \BC^{-1} \left(\BOmega - \bepsilon\right)
- \frac{1}{2}\left(\BA + \BA^T\right) \right] X\, \partial_V \ .
\label{g21}
\end{multline}
Then, from the definition (\ref{g12b}) for $\BA$ and the relation (\ref{g13}), we can show
\begin{equation}
\left[K_X,K_{Q_r}\right] ~=~ \F^T E^{T-1}\, \partial_X \,-\, 
\left[ \dot{\F}^T E - \F^T E^{T-1} \left(\BOmega - \bomega\right) \right] X\,\partial_V \ ,
\label{g22}
\end{equation}
where
\begin{equation}
\F = \dot{F} - \epsilon F \ ,
\label{g23}
\end{equation}
is itself a solution of the oscillator equation. We recognise the r.h.s.~as being of the same form as
$K_{Q_r}$ but with the oscillator solution $F^i{}_{(r)}$ replaced by $\F^i{}_{(r)}$.  This relation therefore 
reproduces the Brinkmann expression (\ref{c17}) for $\left[K_X, K_Q\right]$ for a general
solution $F$. The same analysis also applies of course to the $\left[K_X, K_P\right]$
commutator with the solution $G$.

At this point, it is clear that by choosing the oscillator solutions $F,\, G$ defining $Q_r,\, P_r$
to be the canonical basis $f,\, g$ as in section \ref{sect 3}, we recover the isometry algebra 
(\ref{c5}), (\ref{c6}) precisely. Only the $\left[Q_r,P_s\right]$ commutator has a non-vanishing 
Wronskian, while decomposing the solution $\F$ as in (\ref{c20}) we recover the $\left[X,Q\right]$
commutator from (\ref{g22}), with similar results for $\left[X,P\right]$.

However, it is more constructive to consider the particular choice of generators $Q_a$, $P^a$ 
made above.
We can evaluate the commutators either directly from the expressions for the Killing vectors in 
(\ref{g8}) and (\ref{gg18}), or from the Wronskian forms (\ref{g17}), (\ref{g18}) with oscillator
solutions $F = E$ and $G = E {\cal H}$.
It is straightforward to show:
\begin{align}
\left[K_{Q_a}, K_{Q_b}\right] &= - 2 \bomega_{ab}\, \partial_V~~~=~ - W(E,E)\,\partial_V  \ , 
\label{ggg24} \\
{}&{} \nonumber \\
\left[K_{Q_a}, K_{P^b}\right] &= - \d_a{}^b \, \partial_V  ~~=~ - W(E,E{\cal H})\,\partial_V  \ ,
\label{ggg25} \\
{}&{} \nonumber \\
\left[K_{P^a}, K_{P^b}\right] &=\left( {\cal H} - {\cal H}^T + 2 {\cal H}^T \bomega {\cal H}\right)^{ab} 
~~=~ - W(E{\cal H}, E{\cal H}) \,\partial_V \ .
\label{ggg26}
\end{align}
Using the various identities above for the derivatives of $E^i{}_a(u)$ and ${\cal H}^{ab}(u)$, we can
verify explicitly that the term in brackets in (\ref{ggg26}) is independent of $u$, as it must be
from its representation as a Wronskian. We are therefore free to evaluate it for any value of $u$
and choosing $u=u'$, where ${\cal H}(u') = 0$, we see that it must vanish. We therefore find the 
following extremely simple form for the modified Heisenberg algebra:
\begin{equation}
\left[Q_a, Q_b\right] ~=~ -2 \bomega_{ab} \, Z \ , ~~~~~~~~
\left[P^a, P^b\right] ~=~ 0 \ ,
\label{ggg27}
\end{equation}
and
\begin{equation}
\left[Q_a, P^b\right] ~=~ - \d_a{}^b\, Z \ .
\label{ggg28}
\end{equation}
The $Q_a$ generators therefore develop a non-vanishing commutation relation in the presence
of twist. This is in line with our discussion in section \ref{sect2.3} on the origin of twist. 

It is interesting here to contrast the commutators for $\left[Q_a,Q_b\right]$ and
$\left[P^a,P^b\right]$, the Wronskian for the former being $2\bomega$ and $0$ for the latter.
This seems to reflect the difference in the twist for the two congruences described in 
section \ref{sect 5.2}. The oscillator solution $F = E$ specifying the $Q_a$ generators corresponds 
to (\ref{e8a}) (since there $X ={\rm constant}$) and defines a congruence with twist $\bomega$,
whereas the solution $G = E{\cal H}$ (corresponding to $X = {\cal H} p$) is characteristic of the 
second type of congruence (\ref{e7}) with twist $-\bomega$.

Finally, we need to consider the commutators $\left[X, Q_a\right]$ and $\left[X,P^a\right]$ 
in the special case of homogeneous plane waves. Again, we can evaluate these directly using the
explicit expressions (\ref{g9}), (\ref{gg18}) and (\ref{g11}) for the Killing vectors 
$K_{Q_a}$, $K_{P^a}$ and $K_X$, or alternatively using the general result (\ref{g22}) with the
oscillator solutions $F$ and $G$ defining $Q_a$ and $P^a$. 

Starting with the generator $Q_a$, we easily show that with $F=E$,
\begin{equation}
\F = \dot{F} - \epsilon F = - E \BB \ , ~~~~~~~~~~~~
\dot{\F} = E^{T-1} \left(A - \BOmega B\right) \ ,
\label{ggg29}
\end{equation}
then from (\ref{g22}) we have
\begin{equation}
\left[K_X, K_{Q_a}\right] ~=~ - \BB^T\,\partial_X ~-~ \left(\BA^T + \BB^T \bomega\right) X\,\partial_V \ ,
\label{ggg30}
\end{equation}
as can also be derived directly from the definitions of $K_X$ and $K_{Q_a}$.

The next step is to write this in the form
\begin{equation}
\left[X, Q_a\right] = a_a{}^b Q_b + b_{ab} P^b \ ,
\label{ggg31}
\end{equation}
for constant $a$, $b$.  Comparing the r.h.s.~of (\ref{ggg30}) with the definitions (\ref{g9}), (\ref{gg18})
for $K_Q$ and $K_P$, we require
\begin{equation}
a + b{\cal H}^T = - \BB^T \ , ~~~~~~~~~~~~
b\left( {\bf 1} + 2 {\cal H}^T \bomega\right) = \BA^T \ .
\label{ggg32}
\end{equation}
Differentiating, and using the identities (\ref{ee26}) and (\ref{g14}) for 
$\dot{\cal H}$ and $\dot{\BB}$, $\dot{\BA}$,
we readily find
\begin{equation}
\dot{a} + \dot{b} {\cal H}^T = 0 \ , ~~~~~~~~~~~~ \dot{b} \left({\bf 1} + 2 {\cal H}^T \bomega\right) \ ,
\label{ggg33}
\end{equation}
so conclude that $a$ and $b$ are indeed independent of $u$. Once again evaluating at $u=u'$ where
the terms involving ${\cal H}(u')$ vanish, we therefore determine the commutator
\begin{equation}
\left[X, Q_a\right] ~=~ - \BB^T(u') Q ~+~  \BA^T (u') P \ .
\label{ggg34}
\end{equation}

This just leaves the $\left[X,P\right]$ commutator. A similar calculation shows
\begin{equation}
\G = \dot{G} - \epsilon G = E \left(\dot{\cal H} - \BB {\cal H}\right) \ , ~~~~~~~~~~
\dot{\G} = - E^{T-1} \left[ \left(\BC \BB + \BOmega\right) \dot{\cal H} - 
\left( \BA - \BOmega \BB \right) {\cal H} \right] \ ,
\label{ggg35}
\end{equation}
and from either the analogue of (\ref{g22}) with $\G$, or directly from the definitions of
$K_X$ and $K_{P^a}$, we obtain
\begin{equation}
\left[X,P^a\right] ~=~ \left(\dot{\cal H}^T - {\cal H}^T \BB^T\right) \partial_X
- \left(\dot{\cal H}^T \bomega + \left({\bf 1} + {\cal H}^T \bomega\right) \BB  
+ \tfrac{1}{2} {\cal H}^T \left(\BA + \BA^T\right) \right)X\partial_V \ .
\label{ggg36a}
\end{equation}
This can then be written in the form
\begin{equation}
\left[K_X, K_{P^a}\right]  = c^{ab} Q_b + d^a{}_b P^b \ ,
\label{ggg36}
\end{equation}
with
\begin{equation}
c + d {\cal H}^T = \dot{\cal H}^T - {\cal H}^T \BB^T \ , ~~~~~~~~
d\left({\bf 1} + 2 {\cal H}^T \bomega\right) = \dot{\cal H}^T \left(\BC \BB + 2 \bomega\right) 
+ {\cal H}^T \BA^T \ .
\label{ggg37}
\end{equation}
Then, differentiating (\ref{ggg37}), we can show after some calculation that
\begin{equation}
\dot{c} + \dot{d} {\cal H}^T = 0 \ , ~~~~~~~~~~~~\dot{d} - 2 \dot{c} \bomega = 0 \ ,
\label{ggg38}
\end{equation}
and so verify that $c$ and $d$ are constants.
Evaluating at $u=u'$, we therefore find,
\begin{equation}
\left[X, P^a\right] = \BC^{-1}(u') \, Q + \left( \BB(u') + 2\BC^{-1}(u')  \bomega\right)\,  P \ .
\label{ggg39}
\end{equation}

At this point, we can make the natural consistency check that these commutation relations
satisfy the Jacobi identity.
From (\ref{ggg27}) and (\ref{ggg28}), together with (\ref{ggg34}) and  (\ref{ggg39}),
we have (with all functions evaluated at $u'$),
\begin{align}
&\left[\left[X,Q_a\right],P^b\right] - \left[\left[X, P^b\right],Q_a\right]
+ \left[\left[Q_a,P^b\right],X\right]  \nonumber \\
&=~\left(\BB^T\right)_a{}^b\,Z + 
\left(2 \left(\BC^{-1}\bomega\right)^b{}_a - \left(\BB + 2 \BC^{-1}\bomega\right)^b{}_a \right)\, Z
+ 0\nonumber \\
&=~0 \ .
\label{ggg40}
\end{align}
Notice especially the necessity of the twist appearing in the commutator for $\left[Q_a,Q_b\right]$
of the Heisenberg algebra to ensure the self-consistency of the extended isometry algebra 
for homogeneous plane waves.

The commutators (\ref{ggg34}) and (\ref{ggg39}) may be written out explicitly in terms of
$\BC(u')$, $\BOmega(u')$, $\Bh(u')$ and $\bepsilon(u')$ by simply substituting the definitions
of $\BA$ and $\BB$. The resulting expressions display the dependence
of the commutators $\left[X,Q_a\right]$ and $\left[X,P^a\right]$ on the congruence
defining the Rosen metric through the zweibein $E^i{}_a(u')$ (determining $\BC$, $\Bh$ and $\bepsilon$)
and its derivative $\dot{E}^i{}_a(u')$  (determining $\BOmega$) at the reference point $u'$.
In this explicit form, however, they are quite lengthy and we will not write them here.

In order to complete our self-consistency checks and make contact with the 
canonical set of commutation relations 
described in section \ref{sect3.1}, we may without loss of generality choose $u'=0$
and consider the congruence with $E^i{}_a(0) = \d^i{}_a$ and $\dot{E}^i{}_a(0) = 0$.
Notice though that this is a twist-free congruence, since these conditions imply $\BOmega(0) = 0$
and since $\bomega$ is independent of $u$ it is therefore zero. (Note that this does
not apply to the expansion or shear.)
In this twist-free case, the commutators (\ref{ggg34}) and (\ref{ggg39}) then simplify to
\begin{align}
\left[X,Q_a\right] &= \epsilon_a{}^b\, Q_b + h_{ab}(0) \, P^b \ , \nonumber \\
{}&{} \nonumber \\
\left[X, P^a\right] &= \d^{ab}\, Q_b + \epsilon^a{}_b \, P^b \ , 
\label{ggg41}
\end{align}
reproducing (\ref{c6}).

\subsection{Geodesics and isometries}\label{sect 6.3}

Finally, it is interesting to see how these  $Q_a$ and $P^a$ symmetries  
for a general gravitational plane wave act on the geodesic solutions themselves. 
This discussion follows that given in \cite{Duval:2017els, Zhang:2017rno, Zhang:2017geq} 
in the twist-free case 
with the conventional Rosen metric and geodesics, and we have made this section 
as self-contained as possible to facilitate comparison.
For simplicity, we have also suppressed the index notation below.

Recall the general geodesic solutions from section \ref{sect 5.1},
\begin{equation}
X(u) ~=~ {\cal H}(u) p + a \ , \hskip4cm   ~~~~~~~~~~~
\label{ggg42}
\end{equation}
\begin{equation}
V(u) ~=~ - \frac{1}{2} p {\cal H}(u)p - p {\cal H}^T(u) \bomega a + \eta(u-u') + d \ ,
\label{ggg43}\end{equation}
where $p^a$, $a^a$ and $d$ are constants, $\eta = 0$ for null geodesics,
and the key function ${\cal H}(u)$, which incorporates the twist-dependent path-ordered
exponential ${\cal P}(u,a)$ of (\ref{ee15}), satisfies ${\cal H}(u') = 0$ at the reference point $u'$.

First, under a $Z$ transformation,
\begin{equation}
u \rta u \ , ~~~~~~~~  V \rta V + f \ , ~~~~~~~~ X \rta X \ ,
\label{ggg44}
\end{equation}
the geodesics obviously retain the same form with the simple parameter shift,
\begin{equation}
p \rta p \ , ~~~~~~~~ a \rta a \ , ~~~~~~~~ d \rta d + f \ .
\label{ggg45}
\end{equation}

Next, under the $Q_a$ isometry,
\begin{equation}
u \rta u \ , ~~~~~~~~ V \rta V + c\,\bomega X \ , ~~~~~~~~ X \rta X + c \ ,
\label{ggg46}
\end{equation}
we see that the geodesic solutions transform as
\begin{align}
X(u) &\rta {\cal H}(u) p + a + c \ , \nonumber \\
V(u) &\rta - \frac{1}{2} p {\cal H}(u) p - p {\cal H}^T(u) \left(a + c\right) + \eta(u-u') - a \bomega c \ ,
\label{ggg47}
\end{align}
so again the form of the geodesics is preserved, with the parameter shifts,
\begin{equation}
p \rta p \ , ~~~~~~~~ a \rta a+c \ , ~~~~~~~~ d \rta d - a \bomega c \ .
\label{ggg48}
\end{equation}

Finally, consider the $P^a$ isometry,
\begin{equation}
u \rta u \ , ~~~~~~~~ V \rta V - X\big({\bf 1} - \bomega {\cal H}(u) \big) b 
- \frac{1}{2} b {\cal H}(u) b \ .
\label{ggg49}
\end{equation}
A short calculation shows that here,
\begin{align}
X(u) &\rta {\cal H}(u) (p + b) + a \ , \nonumber \\
V(u) &\rta -\frac{1}{2} (p+b) {\cal H}(u) (p+b) - (p+b) {\cal H}^T(u) \bomega a + \eta (u-u') 
+ d - ab  \nonumber \\
& ~~~~~~ + \frac{1}{2} p \left[ {\cal H} - {\cal H}^T + 2 {\cal H}^T \bomega {\cal H} \right] b \ .
\label{ggg50}
\end{align}
We now recognise the term in square brackets as that occurring in (\ref{ggg26}). 
As explained there, we can verify that its derivative w.r.t.~$u$ vanishes, so that it is independent
of $u$, then evaluating at $u=u'$ with ${\cal H}(u') = 0$ we see that it vanishes.
The form of the geodesics is then once again preserved by the isometry, with the 
parameter transformations,
\begin{equation}
p \rta p + b \ , ~~~~~~~~ a \rta a \ , ~~~~~~~~ d \rta d-ab \ .
\label{ggg51}
\end{equation}

Collecting all this, we therefore find that under a general isometry of the twisted Rosen metric,
the constant parameters specifying the geodesics transform as,
\begin{equation}
\left(p,\,a,\,d,\,\eta\right) ~\rta~ \left(p+b,\,a+c,\,d - a\bomega c - a b + f  \right) \ ,
\label{ggg52}
\end{equation}
generalising the result of \cite{Duval:2017els, Zhang:2017geq} (see eqs.~(III.8), (IV.17) respectively) 
where the corresponding symmetry was identified with a restricted Carroll group.

\section{Van Vleck - Morette Matrix for Twisted Null Congruences}\label{sect 7}

One of the most important geometrical quantities characterising geodesic congruences is the
van Vleck-Morette (VVM) determinant or, more generally, the VVM matrix 
\cite{vanVleck, Pauli, Morette:1951zz,Van Hove}. 
It encodes information on the nature of the geodesic flow and plays a key r\^ole in the construction 
of Green functions and heat kernels for QFTs in curved spacetime (for reviews, 
see {\it e.g.}~\cite{Visser:1992pz, Poisson:2011nh}.
In particular, zeroes of the VVM determinant correspond to conjugate points on the congruence
where the geodesics focus; in turn, this influences the analytic structure of the corresponding
Green functions which is implicitly related to the realisation of causality in the QFT 
\cite{Hollowood:2007ku,Hollowood:2008kq,Hollowood:2009qz,Hollowood:2015elj}.

Here, we generalise the construction of the VVM matrix for plane wave spacetimes previously
given in refs.~\cite{Gibbons:1975jb,Hollowood:2008kq,
  Hollowood:2009qz} 
to the case of null geodesic congruences with twist.
First we give a derivation in terms of the original Brinkmann coordinates then show how the 
result may be obtained directly using the Rosen form (\ref{b18}) of the plane wave metric adapted to
twisted null congruences.

\subsection{Brinkmann construction}

In Brinkmann coordinates, we may define the transverse components of the VVM matrix by

\begin{equation}
\D_{ij}(x,x') = \frac{\partial^2\s(x,x')}{\partial x^i \, \partial x^{\prime j}} \ ,
\label{f1}
\end{equation}
where $\s(x,x')$ is the geodetic interval,
\begin{equation}
\s(x,x') = \frac{1}{2}\int_0^1 d\t\, g_{\m\n} \frac{d x^\m}{d \t} \frac{d x^\n}{d \t}
\label{f2}
\end{equation}
along the geodesic $x^\m(\t)$. For a general plane wave (not necessarily a homogeneous
plane wave as we have been discussing elsewhere), this is
\begin{align}
\s(x,x') &= \frac{1}{2} \int_{u'}^{u} du_1\, 
\left(2 \dot{v} + x^i h_{ij}x^j + \left(\dot{x}^i\right)^2\right) \nonumber \\
&= (u-u') (v-v') + \frac{1}{2}(u-u') \left[x_i \, \dot{x}^i\right]_{u'}^u \ ,
\label{f3}
\end{align}
where we have used the transverse geodesic equation (\ref{b2}) to write
\begin{equation}
\frac{d}{du}\left(x_i \,\dot{x}^i\right) =  x^i h_{ij}x^j + \left(\dot{x}^i\right)^2 \ .
\label{f4}
\end{equation}

As in our earlier work \cite{Hollowood:2008kq, Hollowood:2009qz}, 
we expand the transverse geodesic solutions $x^i(u)$ 
(in other words, the Jacobi fields) as
\begin{equation}
x^i(u) = B^i{}_j(u,u') \,x^j(u') + A^i{}_j(u,u')\, \dot{x}^j(u') \ ,
\label{f5}
\end{equation}
where $A$ and $B$ are solutions of the oscillator/geodesic equation ({\it i.e.}
$\ddot A - hA = 0$, \, $\ddot B - hB = 0$) with ``spray'' and ``parallel'' boundary conditions:
\begin{align}
A^i{}_j(u',u') &= 0, \hskip2cm  \dot{A}^i{}_j(u',u') = \d^i{}_j \ , \nonumber \\
B^i{}_j(u',u') &= \d^i{}_j, \hskip1.8cm  \dot{B}^i{}_j(u',u') = 0 \ , 
\label{f6}
\end{align}
respectively.

These functions are of course closely related to the optical scalars characterising the null congruence.
To see this, recall $\dot{x}^i = \Omega^i{}_j x^j$ and $x^i = E^i{}_a X^a$ so that, suppressing indices,
we have
\begin{align}
x(u) &= B(u,u')\, x(u') + A(u,u') \Omega(u')\, x(u') \nonumber \\
&= E(u) E^{-1}(u') \,x(u') \ .
\label{f7}
\end{align}
Then, from
\begin{equation}
\int_{u'}^u du_1\,  \Omega(u_1) = \int_{u'}^u du_1\,  \dot{E}(u_1) E^{-1}(u_1)
= \left[\,\log E\,\right]_{u'}^u \ ,
\label{f8}
\end{equation}
we find the required relation \cite{Hollowood:2008kq,Hollowood:2009qz}\footnote{As a special case,
if we restrict to a geodesic spray congruence and the corresponding optical scalars, eq.(\ref{f9})
reduces to 
\begin{equation*}
\partial_u \log A(u,u') = \Omega(u) \hskip1cm \Rightarrow \hskip1cm  
\partial_u \log \det A = \tr\, \Omega = \theta \ .
\end{equation*}
For a twist-free congruence, using the relation (\ref{f11}) between $A_{ij}$ and the VVM matrix $\D_{ij}$,
this directly implies the well-known relation \cite{Visser:1992pz} 
between the VVM determinant and the expansion scalar:
\begin{equation*}
\det \D(u,u') = (u-u')^2 \, e^{-\int_{u'}^u du^\dprime\, \theta(u^\dprime) }\ .
\end{equation*}}
\begin{align}
B(u,u') + A(u,u') \Omega(u') &= \exp \int_{u'}^u du_1\, \Omega(u_1) \nonumber \\
&= \exp \int_{u'}^u du_1\, \left[ \frac{1}{2} \theta {\bf 1} + \s + \omega \right] \ .
\label{f9}
\end{align}

Now, we can express the geodetic interval in terms of $A(u,u')$ and $B(u,u')$ as follows:
\begin{align}
\s(x,x') = (u-u')(v-v') - \frac{1}{2}(u-u')\, &\left[\, x(u) 
\left(A^{-1,T}(u,u') - A^{-1}(u',u)\right)x(u') \right.\nonumber \\
&+ x(u) \left(A^{-1}(u',u) B(u',u) \right) x(u) \nonumber \\
&\left.- x(u') \left(A^{-1}(u,u') B(u,u')\right) x(u') \, \right]\ ,
\label{f10}
\end{align}
and so the VVM matrix is 
\begin{equation}
\D_{ij}(u,u') = - \frac{1}{2} (u-u') \,\left[\,A^{-1,T}(u,u') - A^{-1}(u',u)\,\right] \ .
\label{f11}
\end{equation}
Notice immediately though that we can not assume that $A^T(u,u') = - A(u',u)$ in the case of a 
congruence with twist, unlike the conventional twist-free case 
described in \cite{Hollowood:2008kq,Hollowood:2009qz}.

Determining the VVM matrix therefore reduces to finding the solution $A(u,u')$ of the 
oscillator/geodesic equation with spray boundary conditions. We already have one solution 
in the form of the zweibein $E^i{}_a(u)$, so we can find a second solution using the
Wronskian, {\it i.e.}
\begin{equation}
E^T(u) \dot{A}(u,u') - \dot{E}^T(u) A(u,u') = W(u') \ ,
\label{f12}
\end{equation}
where $W_{aj}(u')$ is independent of $u$ since the oscillator equation is second order with no terms
linear in derivatives. 
Now rearrange to find a differential equation for $E^{-1}A$. First, from (\ref{f12}) we have
\begin{equation}
E^{-1} \dot{A} = E^{-1}\Omega^T A + \BC^{-1} W
\label{f13}
\end{equation}
recalling $\BC = E^T E$. Expressing $\Omega^T = \Omega - 2\omega$, some further
manipulation then gives
\begin{equation}
\frac{d}{du}\left(E^{-1} A\right) +  2 \BC^{-1} \bomega \left(E^{-1} A\right) = \BC^{-1} W \ ,
\label{f14}
\end{equation}
in terms of the Rosen twist $\bomega_{ab}$.

To solve this, we need the path-ordered exponential ${\cal P}(u,a)$ introduced in (\ref{ee15}).
Since this is an integrating factor for the differential equation (\ref{f14}), the general solution is 
\begin{equation}
{\cal P}(u,a) E^{-1}(u) A(u,u') - {\cal P}(u^\dprime,a) E^{-1}(u^\dprime) A(u^\dprime,u') = 
\int_{u^\dprime}^u du_1 \, {\cal P}(u_1,a) \,\BC^{-1}(u_1) \, W \ .
\label{f16a}
\end{equation}
Choosing $u^\dprime=u'$ and using the boundary conditions (\ref{f6}) to set $A(u',u') = 0$,
then taking $a=u$, we find the solution in the convenient form,
\begin{equation} 
E^{-1}(u) A(u,u') = \int_{u'}^u du_1\, {\cal P}(u_1,u)\, \BC^{-1}(u_1) \,W \ .
\label{f16}
\end{equation}
Now, since $W$ is independent of $u'$, we may evaluate its definition (\ref{f12}) at $u=u'$ to find
\begin{equation}
W(u') = E^T(u') \dot{A}(u',u') - \dot{E}^T(u') A(u',u') = E^T(u') \ ,
\label{f17}
\end{equation}
again imposing the boundary conditions (\ref{f6}). 

Finally, therefore, we find
\begin{equation}
A(u,u') = E(u) \, \int_{u'}^u du_1 \, {\cal P}(u_1,u) \, \BC^{-1}(u_1) \, E^T(u') \ ,
\label{f18}
\end{equation}
or more explicitly,
\begin{equation}
A(u,u') = E(u) \, \left(\, \int_{u'}^u du_1 \, {\cal T}_+\exp\left[-\int_{u_1}^u dt 
\,2\BC^{-1}(t) \bomega\right]\, \BC^{-1}(u_1) \,\right)\, E^T(u') \ .
\label{f19}
\end{equation}
The difference from our previous result \cite{Hollowood:2008kq, Hollowood:2009qz} 
for twist-free congruences is clearly the inclusion of the twist-dependent integrating factor.
The VVM matrix then follows from (\ref{f11}).

\subsection{Rosen construction}

In Rosen coordinates, we define the transverse VVM matrix $\BDelta_{ab}(u,u')$ in terms of the geodetic interval
as
\begin{equation}
\BDelta_{ab}(u,u') = E_{\,a}^{T\,i}(u) \D_{ij}(u,u') E^j{}_b(u') = \frac{\partial^2\s(x,x')}{\partial X^a \,\partial X^{\prime\, b} }\ .
\label{f20}
\end{equation}
From the Rosen metric (\ref{b18}) or (\ref{e1}), we have
\begin{align}
\s(x,x') &= \frac{1}{2}(u-u') \, \int_{u'}^u du_1\,\left(2\dot{V} - 2 X^a \bomega_{ab}\dot{X}^b 
+ \dot{X}^a \BC_{ab} \dot{X}^b  \right) \nonumber \\
&= (u-u') (V-V') + \frac{1}{2}(u-u') \left[X \left(\BC\dot{X} + 2 \bomega X\right) \right]_{u'}^u \ ,
\label{f21}
\end{align}
where, assuming $\bomega$ independent of $u$ as in (\ref{b19}) to maintain the equivalence with the 
Brinkmann congruence, we have used the transverse geodesic equation (\ref{e4}),
\begin{equation}
\frac{d}{du}\,\left(\BC \dot{X} + 2 \bomega X\right) = 0 \ .
\label{f22}
\end{equation}
which implies (\ref{e5a}),
\begin{equation}
\dot{X} + 2 \BC^{-1} \bomega X = \BC^{-1} \xi \ ,
\label{f23}
\end{equation}
for constant $\xi^a$. As discussed in section \ref{sect 5}, the solution is 
\begin{equation}
{\cal P}(u,a)\, X(u) - {\cal P}(u',a)\, X(u') =
\int_{u'}^u du_1 \, {\cal P}(u_1,a)\, \BC^{-1}(u_1) \, \xi \ .
\label{f26}
\end{equation}
Inverting now gives
\begin{equation}
\xi = \left(\,\int_{u'}^u du_1 \, {\cal P}(u_1,a)\, \BC^{-1}(u_1)\,\right)^{-1} \,
\Big[\,{\cal P}(u,a)\, X(u) - {\cal P}(u',a)\, X(u')\,\Big] \ .
\label{f27}
\end{equation}

It then follows from the expression (\ref{f21}) for the geodetic
interval that\footnote{Note that the presence of the twist-dependent integrating factor in (\ref{f28})
means that the geodetic interval is not a simple quadratic form in $(X-X')$, in contrast to the
twist-free case \cite{Hollowood:2008kq, Hollowood:2009qz} where
\begin{equation*}
\s(x,x') = (u-u')(V-V') + \frac{1}{2} (X-X')^a \D_{ab}(u,u') (X-X')^b \ .
\end{equation*}  }
\begin{align}
\s(x,x') &= (u-u') (V-V') \nonumber \\
&+ \frac{1}{2} (u-u') (X-X')\,\left(\,\int_{u'}^u du_1 \, {\cal P}(u_1,a)\, \BC^{-1}(u_1)\,\right)^{-1} \,
\nonumber \\
&\hskip6cm \times \Big[\,{\cal P}(u,a)\, X(u) - {\cal P}(u',a)\, X(u')\,\Big] \ .
\label{f28}
\end{align}
From its definition (\ref{f20}), the VVM matrix is then
\begin{equation}
\BDelta_{ab}(u,u') = - \frac{1}{2} (u-u')\,\Big[\, \BA^{-1,T}(u,u') - \BA^{-1}(u,u')\,\Big] \ ,
\label{f29}
\end{equation}
where
\begin{equation}
\BA^{-1}(u',u) = - \left(\,\int_{u'}^u du_1 \, {\cal P}(u_1,a)\, \BC^{-1}(u_1)\,\right)^{-1} \, {\cal P}(u',a) \ .
\label{f30}
\end{equation}
That is, combining the integrals, or equivalently choosing $a = u'$, 
\begin{align}
\BA(u,u') &= \int_{u'}^u du_1\,{\cal P}(u_1,u')\,\BC^{-1}(u_1) \nonumber \\
&\equiv \int_{u'}^u du_1\,  
{\cal T}_+ \exp\left[-\int_{u_1}^u dt \,2\BC^{-1}(t) \bomega\right] \,\BC^{-1}(u_1)  \ .
\label{f31}
\end{align}
Finally, lowering indices with the Rosen metric and with $\BA(u,u') = E^T(u) A(u,u') E(u')$, 
we recover the Brinkmann form of the VVM matrix derived above.

\section{Discussion}\label{sect 8}

In this paper, we have established the mathematical framework to describe the geometry of twisted
null congruences in gravitational plane wave spacetimes, with a special focus on homogeneous 
plane waves. 

Since these metrics arise as Penrose limits, our results are of sufficient generality 
to encompass any application where the essential physics is governed by the geometry of 
geodesic deviation. Notably, this includes loop effects in quantum field theory
in general curved spacetimes, with applications ranging from ultra-high
energy particle scattering to the origin of matter-antimatter asymmetry.

Existing studies of quantum field propagation in gravitational backgrounds have almost 
entirely been restricted to the conventional geometric optics description associated with 
plane waves, where the classical null rays form a gradient flow. 
That is, the corresponding null geodesic congruence exhibits only expansion and shear.
While little is known at present about the nature of quantum field theoretic effects associated 
with twisted null congruences (for some related classical studies, 
see {\it e.g.}~\cite{Nurowski:1992yn, Newman:2004ba, Davidson}),
our aim here has been to develop a comprehensive geometric toolkit to enable future
work in this area.

Moreover, the importance of gravitational plane waves as string backgrounds in itself motivates 
the most intensive exploration of the geometry of these spacetimes. It may be hoped that 
our novel description of the r\^ole of twist in this geometry may find useful applications
in string theory itself.

A key focus of our work was the relation of Rosen coordinates, which reflect the nature of
a chosen null congruence, to the more fundamental Brinkmann coordinates,
and we derived the generalised Rosen metric (\ref{b18}) adapted to a twisted congruence.
The modifications due to twist of many geometrical constructions relevant to 
loop calculations in QFT in curved spacetime were discussed in detail, notably the
generalised form of the van Vleck-Morette matrix and its Brinkmann-Rosen correspondence,
and a thorough description of isometries of (homogeneous) plane waves in both the 
Brinkmann and twisted Rosen descriptions was presented.

Geodesic deviation is also central to the detection of gravitational waves in an astronomical 
context. At the most basic level, the passage of a gravitational wave may be detected 
by its effect on a ring of freely-falling test particles. This
``Tissot circle'' \cite{Zhang:2017rno, Zhang:2017geq}
is simply a cross-section of a (timelike) geodesic congruence, and the squeezing and squashing
measured by the detector, for example the arms of an interferometer such as
LIGO \cite{Abbott:2016blz, Abbott:2016nmj} or eLISA \cite{AmaroSeoane:2012je}, 
is the expansion and shear of the congruence in the gravitational wave metric.
The geometric results presented here would therefore be relevant if a detector would
in addition be sensitive to a rotation of the particles on the Tissot circle, {\it i.e.}~the
twist of the corresponding geodesic congruence. In principle, recent proposals 
\cite{Loeb:2015ffa, Kolkowitz:2016wyg}
to detect gravitational waves through measurements of the induced effective
Doppler shifts on ultra-precise optical lattice atomic clocks in space
may ultimately allow this possibility to be realised. 

These astrophysical gravitational waves would be in the form of short duration bursts
or pulses, in which the plane wave profile function $h_{ij}(u)$ vanishes outside a
given range. As discussed recently in \cite{Zhang:2017rno, Zhang:2017geq} 
(see also references therein) 
the shape of this profile function encodes information on the nature of the source,
with its iterated integrals, in both Brinkmann and Rosen coordinates, distinguishing
different phenomena such as the memory effect.

In conclusion, this geometry of twisted null congruences in (homogeneous) gravitational 
plane waves is mathematically elegant and a natural extension of existing results in this 
important area of general relativity. Our hope is that it will provide an impetus to further
explorations of the physics of twist in quantum field theory and string theory
in curved spacetime as well as for gravitational plane waves in astronomy.

\vskip0.7cm
\noindent{\bf Acknowledgments}

\noindent I am grateful to Tim Hollowood for many discussions and
collaboration on plane wave geometry, and Gary Gibbons for bringing
refs.\cite{Duval:2017els,Zhang:2017rno,Zhang:2017geq}
to my attention.
This work was supported in part by STFC grant ST/L000369/1.

\vskip1cm

\appendix

\section{Null Congruences in an anti-Mach Spacetime}\label{sect 4}

In this appendix, we study in detail the null geodesics and congruences in the homogeneous plane wave
metric \ref{c1}. We describe explicit solutions to the geodesic equations for this generalised 
Ozsv{\'a}th-Sch{\"u}cking (OS) metric in different coordinate systems, including a
Newman-Penrose basis, and discuss the Raychoudhuri equations and optical scalars 
for a twisted null congruence.

\subsection{Generalised Ozsv{\'a}th-Sch{\"u}cking metric and co-rotating coordinates}\label{sect 4.1}

We can write the generalised OS metric in the alternative forms,
\begin{align}
ds^2 &= 2 du dv + \left(e^{\e u} h_0 e^{-\e u}\right)_{ij} x^i x^j du^2 + \left(dx^i\right)^2 \nonumber \\
&= 2 du dv + x\, O(u) h_0 O^T(u)\, x\, du^2 + \left(dx^i\right)^2 \ ,
\label{d1}
\end{align}
where 
\begin{equation}
O(u) = e^{\e u} = \begin{pmatrix}\cos u &\,\, \sin u \\ -\sin u &\,\, \cos u \end{pmatrix} \ ,
\label{d2}
\end{equation}
is an orthogonal matrix. We will use these two representations interchangeably in what follows, depending
on which is most transparent at a given step.

We consider first an arbitrary profile function $h_0 = \begin{pmatrix} a &\, 0 \\ 0 &\, b\end{pmatrix}$
with $a$, $b$ constant. The non-vanishing Riemann, Ricci and Weyl curvature components are
\begin{align}
R_{uiuj} &= - h_{ij} \ , \hskip2cm  R_{uu} = - \tr\, h_0 = -(a+b) \ , \nonumber \\
{}&{} \nonumber \\
C_{uiuj} & = -\tfrac{1}{2} (a-b) \begin{pmatrix}\cos 2u & \,\, -\sin 2u \\ -\sin 2u &\,\,  -\cos 2u \end{pmatrix} \ ,
\label{d3}
\end{align}
where $R_{uiuj} = C_{uiuj} + \tfrac{1}{2}R_{uu} \d_{ij}$.
From (\ref{b25}), (\ref{b26}) we then have the NP curvature scalars in the standard basis,
\begin{equation}
\Phi_{22} = \tfrac{1}{2} (a+b) \ , \hskip2cm
\Psi_4 = \tfrac{1}{2}(a-b) e^{2iu} \ .
\label{d3a}
\end{equation}
Evidently the metric with $a=b$ is conformally flat, while $a=-b$ gives a Ricci flat spacetime.
The original OS, anti-Mach metric is the Ricci-flat solution 
with $a=1$, $b=-1$.

It is clear from \ref{d1} that a natural choice of transverse coordinates would be to take out the
rotation in the profile function and define
\begin{equation}
z^i = \left(e^{-\e u} x\right)^i = O^T(u)\, x  \ .
\label{d4}
\end{equation}
The metric in these co-rotating, or stationary, coordinates becomes 
\begin{align}
ds^2 &= 2 du dv + \left( h_0 + \bf{1}\right)_{ij} z^i z^j \, du^2  - 2 \e_{ij} z^i dz^j \, du 
+ \left(dz^i\right)^2  \nonumber \\
&= 2 du dv + \left( (a+1) (z^1)^2 + (b+1)(z^2)^2 \right) du^2 + 2\left(z^2 dz^1 - z^1 dz^2\right) du 
+ \left(dz^i\right)^2  \ .
\label{d5}
\end{align}

The curvature tensors are especially simple in these coordinates.
The Riemann and Weyl tensors are 
\begin{equation*}
R_{uiuj} = - \left(h_0\right)_{ij} = - \begin{pmatrix} a&\,0 \\ 0&\,b \end{pmatrix} \ , \hskip 2cm
C_{uiuj} = -\tfrac{1}{2} (a-b) \begin{pmatrix}1&\,0 \\0 &\,-1 \end{pmatrix} \ ,
\label{d5a}
\end{equation*}
while of course $R_{uu} = -(a+b)$ as before.  The $X$ isometry also simplifies. 
The manifest invariance of the metric (\ref{d5}) under translations
$u \rightarrow u+\a$ implies the Killing vector is now just $K_X = \partial_u$, the accompanying 
rotation on the $x^i$ coordinate in (\ref{c13}) being removed
by the transformation to co-rotating coordinates.\footnote{For the Ricci-flat, Ozsv{\'a}th-Sch{\"u}cking metric
$a=1$, $b=-1$, a further coordinate transformation $U= \sqrt{2}\, u$, 
$W = \tfrac{1}{\sqrt{2}}\,\left(v + z^1 z^2\right)$ brings the metric to 
\begin{equation*}
ds^2 = 2 dU dW -2\sqrt{2} \,z^1 dz^2 \,dU + (z^1)^2\, dU^2 + \left(dz^i\right)^2  \ ,
\end{equation*}
which, up to normalisation factors, is the form given in the original OS paper \cite{OS}.
Evidently, this has 3 commuting isometries corresponding to translations in $U$, $W$ and $z^2$.
In terms of the generators in (\ref{c5}) and (\ref{c6}), these are $X$, $Z$ and the linear combination 
$Q_2 + P_1$, which for $b=-1$ commutes with both $Z$ and $X$ \cite{Blau:2002js}. }

The null geodesic equations, written explicitly in terms of the co-rotating coordinates, are\footnote{
In the co-rotating transverse coordinates $z^i$, the Christoffel symbols are
\begin{equation*}
\Gamma^v_{uu} = (a-b) z^1 z^2 \ , \hskip1cm
\Gamma^v_{ui} = (h_0)_{ij} z^j \ , \hskip1cm
\Gamma^i_{uu} = -\left(h_0 + {\bf 1}\right)_{ij} z^j \ , \hskip1cm
\Gamma^i_{uj} = \e^i{}_j \ .
\end{equation*}\label{fn2}}
\begin{align}
&\ddot{v} + 2a z^1 \dot{z}^1 + 2b z^2 \dot{z}^2 + (a-b)z^1 z^2 = 0 \ ,\nonumber \\
&\ddot{z}^1 + 2 \dot{z}^2 - (a+1) z^1 = 0 \ ,\nonumber \\
&\ddot{z}^2 - 2 \dot{z}^1 - (b+1) z^2 = 0 \ .
\label{d12}
\end{align}

It is also useful to note the explicit form of the standard Newman-Penrose tetrad in these coordinates.
A straightforward construction starting from $\ell_\m = \partial_\m u$ gives
\begin{equation}
\ell^\m = \begin{pmatrix}0\\1\\{\bf 0}\end{pmatrix} \ , \hskip1.5cm
n^\m = \begin{pmatrix}-1\\\tfrac{1}{2}z(h_0 + {\bf 1})z\\{\bf 0}\end{pmatrix} \ , \hskip1.5cm
m^\m = \frac{1}{\sqrt2} \begin{pmatrix} 0\\-z^2 + i z^1\\\d^{i1} + i \d^{i2}\end{pmatrix}
\label{d6}
\end{equation}
The corresponding NP curvature scalars are then simply 
\begin{equation}
\Phi_{22} = - \tfrac{1}{2}R_{\m\n}n^\m n^\n = \tfrac{1}{2} {\rm tr}\,h_0 = \tfrac{1}{2}(a+b) \ , \hskip1.5cm
\Psi_4 = - C_{n\bar{m}n\bar{m}} = \tfrac{1}{2}(a-b) \ .
\label{d7}
\end{equation}

\subsection{Oscillator solutions}

In the original coordinates, the null geodesic equations are given in (\ref{b2}). Importantly, the equations 
for the transverse coordinates $x^i$ are solutions of the oscillator equation and we focus on these. 
That is, we look for explicit solutions of the oscillator equation,
\begin{equation}
\ddot{F}^i = h^i{}_j(u) F^j \ ,
\label{d8}
\end{equation}
with
\begin{equation}
h = e^{\e u} \begin{pmatrix} a &\,0\\0&\,b\end{pmatrix}e^{-\e u} = 
O(u) \begin{pmatrix} a &\,0\\0&\,b\end{pmatrix} O^T(u) \ .
\label{d9}
\end{equation}
We can immediately write these solutions in the form (suppressing indices)
\begin{equation}
F = e^{\e u} {\textsl f} = O(u)  {\textsl f} \ ,
\label{d10}
\end{equation}
where 
\begin{equation}
\ddot{{\textsl f}\,\,} + 2\e \dot{{\textsl f}\,\,} - (h_0 + {\bf 1}) {\textsl f} = 0 \ .
\label{d11}
\end{equation}
This corresponds to the geodesic equations (\ref{d12}) for the co-rotating transverse coordinates.

To solve these equations, we first make the ansatz
\begin{equation}
{\textsl f} = P e^{\e\l u} X = P O(\l u) X  \ ,
\label{d13}
\end{equation}
where $X$ is a constant vector, $P = \begin{pmatrix}\a&\, 0\\0&\,\,\b\end{pmatrix}$
and $\a$, $\b$, $\l$ are to be determined.
Without loss of generality, we can immediately rescale so that $\a = 1$. Substituting in (\ref{d11}), 
we require
\begin{align}
\ddot{{\textsl f}\,\,} + 2\e \dot{{\textsl f}\,\,} - (h_0 + {\bf 1}) {\textsl f} 
&~=~ \left[ -\l^2 + 2\l \e P \e P^{-1} - \left(h_0 + {\bf 1}\right)\right] {\textsl f} \nonumber \\
&~=~ 0 \ ,
\label{d14}
\end{align}
so we find a solution if $\b$, $\l$ satisfy\footnote{It will be useful in later calculations
to eliminate $\b$ to obtain $\l$ directly as a solution of
\begin{equation*}
\l^4 + (a+b-2)\l^2 + (a+1)(b+1) = 0 \ .
\end{equation*}
\label{fn3}}
\begin{align}
\l^2 + 2\b \l + a + 1 &= 0 \ , \nonumber \\
\l^2 + \frac{2}{\b}\l + b + 1 &= 0 \ .
\label{d15}
\end{align}
In the OS ($a=1$, $b=-1$) metric, we have $\b = \pm\sqrt{2}$ and
$\l = \mp\sqrt{2}$.

The remaining two solutions are not so simple in general, but if we specialise to the 
OS metric we can show that
\begin{equation}
{\textsl f} = \begin{pmatrix} 1 &\,0 \\u &\, 1\end{pmatrix} X \ ,
\label{d16}
\end{equation}
with $X$ a constant vector, also solves (\ref{d11}).

To summarise, in the most interesting case of the Ricci-flat, OS spacetime, we have a complete set of
solutions to the oscillator equation:
\begin{align}
{\textsl f}^{\,\, i}_{(1)} &~=~ O(u) \begin{pmatrix}1&\,0\\ 0 &\, -\sqrt2 \end{pmatrix} O(\sqrt{2} u) 
\begin{pmatrix}1\\0\end{pmatrix} \ , \nonumber \\
{\textsl f}^{\,\, i}_{(2)} &~=~ O(u) \begin{pmatrix}1&\,0\\ 0 &\, -\sqrt2 \end{pmatrix} O(\sqrt{2} u) 
\begin{pmatrix}0\\1\end{pmatrix} \ , \nonumber \\
{\textsl f}^{\,\, i}_{(3)} &~=~ O(u) \begin{pmatrix}1&\,0\\ u &\, 1 \end{pmatrix} 
\begin{pmatrix}1\\0\end{pmatrix}  ~=~ O(u) \begin{pmatrix}1\\u \end{pmatrix} \ , \nonumber \\
{\textsl f}^{\,\, i}_{(4)} &~=~ O(u) \begin{pmatrix}1&\,0\\ u &\, 1 \end{pmatrix} 
\begin{pmatrix}0\\1\end{pmatrix}  ~=~ O(u) \begin{pmatrix}0\\1 \end{pmatrix} \ .
\label{d17}
\end{align}
The solutions $f^i_{(r)}$, $g^i_{(r)}$, $r = 1,2$ introduced in section \ref{sect2.3} are linear combinations
of these solutions chosen to satisfy the canonical boundary conditions (\ref{b20}). They may be compared
directly with the solutions (3.40) of ref.~\cite{Blau:2002js}.

\subsection{Twisted null congruence and optical scalars}\label{sect 4.3}

As discussed in section \ref{sect2.3}, to select a null congruence with the geodesic equations satisfied 
by transverse coordinates $x^i = E^i{}_a(u) X^a$, we choose {\it half} of the oscillator solutions 
to form the zweibein $E^i{}_a$. Here, instead of the canonical choice, we study the natural
congruence picked out by the first two solutions above and define $E^i{}_a = F^i{}_{(a)}$, $a=1,2$.
This implies
\begin{equation}
x^i = E^i{}_a X^a = O(u) \begin{pmatrix}1 &\,0\\ 0&\,\b\end{pmatrix} O(\l u) 
\begin{pmatrix}X^1 \\ X^2\end{pmatrix} \ ,
\label{d18}
\end{equation}
for the general metric, with the integration constants $X^a$ interpreted as Rosen coordinates.

The corresponding Wronskian is therefore (recalling that we may evaluate 
at $u=0$ because $E^i{}_a$ satisfy the oscillator equation)
\begin{align}
W_{ab} &= \left(E^T \dot{E}\right)_{ab} - \left(\dot{E}^T E\right)_{ab} \nonumber \\
&= \left(2\b + \left(\b^2 + 1\right)\l \right) \e_{ab} \ .
\label{d19}
\end{align}
This null congruence therefore has a non-vanishing twist, 
\begin{equation}
\bomega_{ab} = \left(\b + \tfrac{1}{2}\left(\b^2 + 1\right)\l \right) \e_{ab} \ .
\label{d20}
\end{equation}
in Rosen coordinates. The corresponding Brinkmann twist is
\begin{align}
\omega_{ij} &= \left((E^T)^{-1} \bomega E^{-1}\right)_{ij} \nonumber \\
&= \left(\b + \tfrac{1}{2}\left(\b^2 + 1\right)\l \right) O(u) \begin{pmatrix}1&\,0\\0&\,1/\b\end{pmatrix} 
\e \begin{pmatrix}1&\,0\\0&\,1/\b\end{pmatrix} O^T(u) \nonumber \\
&= \L \,\e_{ij} \ ,
\label{d21}
\end{align}
where $\L = \left(1+ \tfrac{1}{2}\left(\b + 1/\b\right)\l \right)$. For the OS metric, $\L = -1/2$.

We can construct the full set of optical scalars directly from the general formulae in section \ref{sect2.1}.
From the definition (\ref{b4}), we find
\begin{align}
\Omega_{ij} &= \left(\dot{E} E^{-1}\right)_{ij} \nonumber \\
&= O(u) \begin{pmatrix} 0&\,\, 1 + \l/\b \\ -(1+\l\b) &\,\, 0 \end{pmatrix} O^T(u) \ ,
\label{d22}
\end{align}
then read off the optical scalars from the decomposition (\ref{b10}). Evidently $\Omega_{ij}$ is traceless
so the expansion scalar $\theta$ vanishes. Its symmetric part gives the shear,
and we check that the antisymmetric part reproduces the twist (\ref{d19}).
Simplifying the resulting expressions using the defining equations (\ref{d15}) for $\b$ and $\l$, 
we eventually find,
\begin{equation}
\theta = 0 \ , \hskip1.5cm 
\s_{ij} = \tfrac{1}{4} (a-b) \begin{pmatrix} \sin 2u &\,\, \cos 2u \\\cos 2u &\,\, -\sin 2u \end{pmatrix} \ ,
\hskip1.5cm  \omega_{ij} = \L\, \e_{ij} \ ,
\label{d222}
\end{equation}
with $\L = \tfrac{1}{2}\left(1 - \l^2 - \tfrac{1}{2}(a+b)\right)$.

The Raychaudhuri equations (\ref{b12}) simplify as a result of the vanishing of $\theta$. We immediately
have 
\begin{align}
\dot{\theta} &= 0 \ , \nonumber \\
\dot{\s}_{ij} &= \tfrac{1}{2}(a-b) \begin{pmatrix} \cos 2u &\,\, -\sin 2u \\-\sin 2u &\,\, -\cos 2u \end{pmatrix}
= - C_{uiuj} \ , \nonumber \\
\dot{\omega}_{ij} &= 0 \ ,
\label{d23}
\end{align}
where for the first we need to verify  $\tr\, \s^2 + \tr\, \w^2 = - R_{uu} = (a+b)$, which follows 
using the identity in footnote \ref{fn3}.

\subsection{Null geodesics}

Through the Raychaudhuri equations, the solutions of the geodesic equations for the transverse coordinates 
control the essential features of the null congruence. Now, we want to focus on the properties of an 
individual null geodesic, so we also require the solution for the coordinate $v$. This discussion is best
made in terms of the co-rotating coordinates, so we start from the set of geodesic equations (\ref{d12}).

First note that a first integral of the geodesic equation for $v$ follows immediately from imposing
the null condition on the metric (\ref{d5}), giving
\begin{equation}
2 \dot{v} + z^i (h_0 + {\bf 1})_{ij} z^j - 2 \e_{ij} z^i \dot{z}^j  + \left(\dot{z}^i \right)^2  = 0 \ .
\label{d24}
\end{equation}
The transverse geodesic equations can be written in compact form as
\begin{equation}
\ddot{z}^i + 2 \e^i{}_j \dot{z}^j - (h_0 + {\bf 1})_{ij} z^j = 0 \ .
\label{d25}
\end{equation}
The geodesics (\ref{d18}) forming the twisted null congruence are given by
\begin{equation}
z = \begin{pmatrix}1 &\, 0 \\0 &\,\b\end{pmatrix} O(\l u) X \ ,
\label{d26}
\end{equation}
which implies
\begin{equation}
\dot{z} = \l \begin{pmatrix}1 &\, 1/\b \\-\b &\,0\end{pmatrix} z \ , \hskip2cm
\ddot{z} = -\l^2 z \ .
\label{d27}
\end{equation}
A short calculation gives the consistency check
\begin{equation}
\ddot{z} + 2 \e \dot{z} - (h_0 + {\bf 1}) z = 
- \begin{pmatrix}\l^2 + 2\b\l + a+1 &\,\,0\\0 &\,\, \l^2 + \tfrac{2}{\b}\l +b+1\end{pmatrix} z = 0 \ ,
\label{d28}
\end{equation}
by virtue of the equations (\ref{d15}) defining $\b$ and $\l$.

Next, substituting the explicit solution for $\dot{z}$ into (\ref{d24}) gives
\begin{align}
\dot{v} &= - \tfrac{1}{2}\left(\b^2 - 1\right) \l^2 ~ 
z \begin{pmatrix}1 &\,\, 0 \\ 0 &\,\, - 1/\b^2\end{pmatrix} z \nonumber \\
&=  - \tfrac{1}{2}\left(\b^2 - 1\right) \l^2 ~
X O^T(\l u) \begin{pmatrix}1 &\, 0 \\ 0 &\, -1\end{pmatrix}O(\l u) X \ .
\label{d29}
\end{align}
Collecting earlier results, we can verify that this is consistent with the original form (\ref{d3}) for $v$,
which implies
\begin{equation}
\dot{v} = -\tfrac{1}{2} X^a \dot{\bf\Omega}_{ab} X^b = 
-\tfrac{1}{2} X E^T \left(h + \Omega^T \Omega\right) E X \ .
\label{d30}
\end{equation}

Integrating to find $v$ itself, and writing out the solutions explicitly, we finally find
\begin{align}
v &= V -\tfrac{1}{4}\left(\b^2 -1\right) \l ~
X \begin{pmatrix} \sin 2\l u &\,\, \cos 2\l u \\ \cos 2\l u &\,\, -\sin 2\l u \end{pmatrix} X \ ,
\nonumber \\
z&= \begin{pmatrix} \cos\l u &\,\, \sin\l u \\ -\b \sin\l u &\,\, \b \cos \l u \end{pmatrix} X \ .
\label{d31}
\end{align}

\begin{figure}[h!]
\centering
\includegraphics[scale=0.6
]{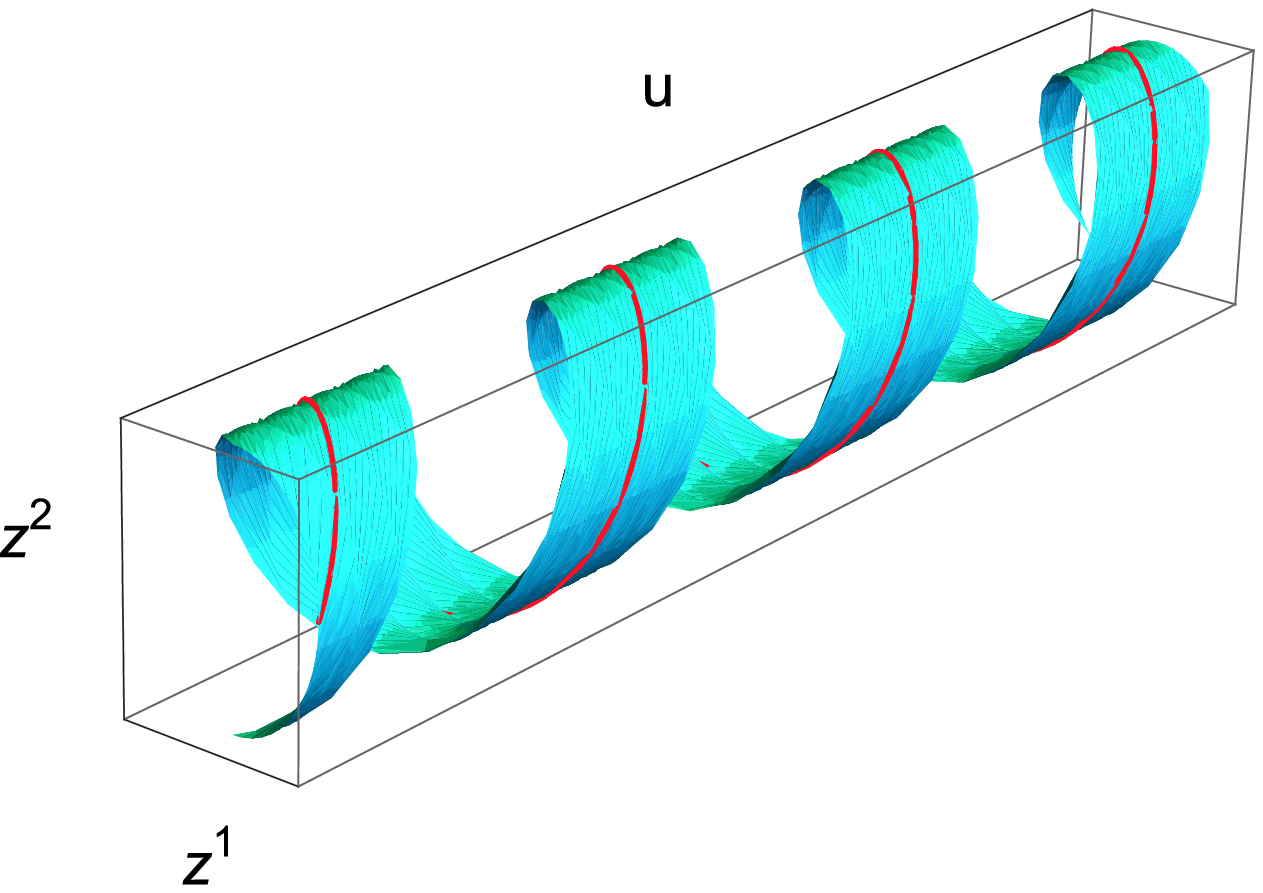} \hskip0.3cm
\includegraphics[scale=0.5]{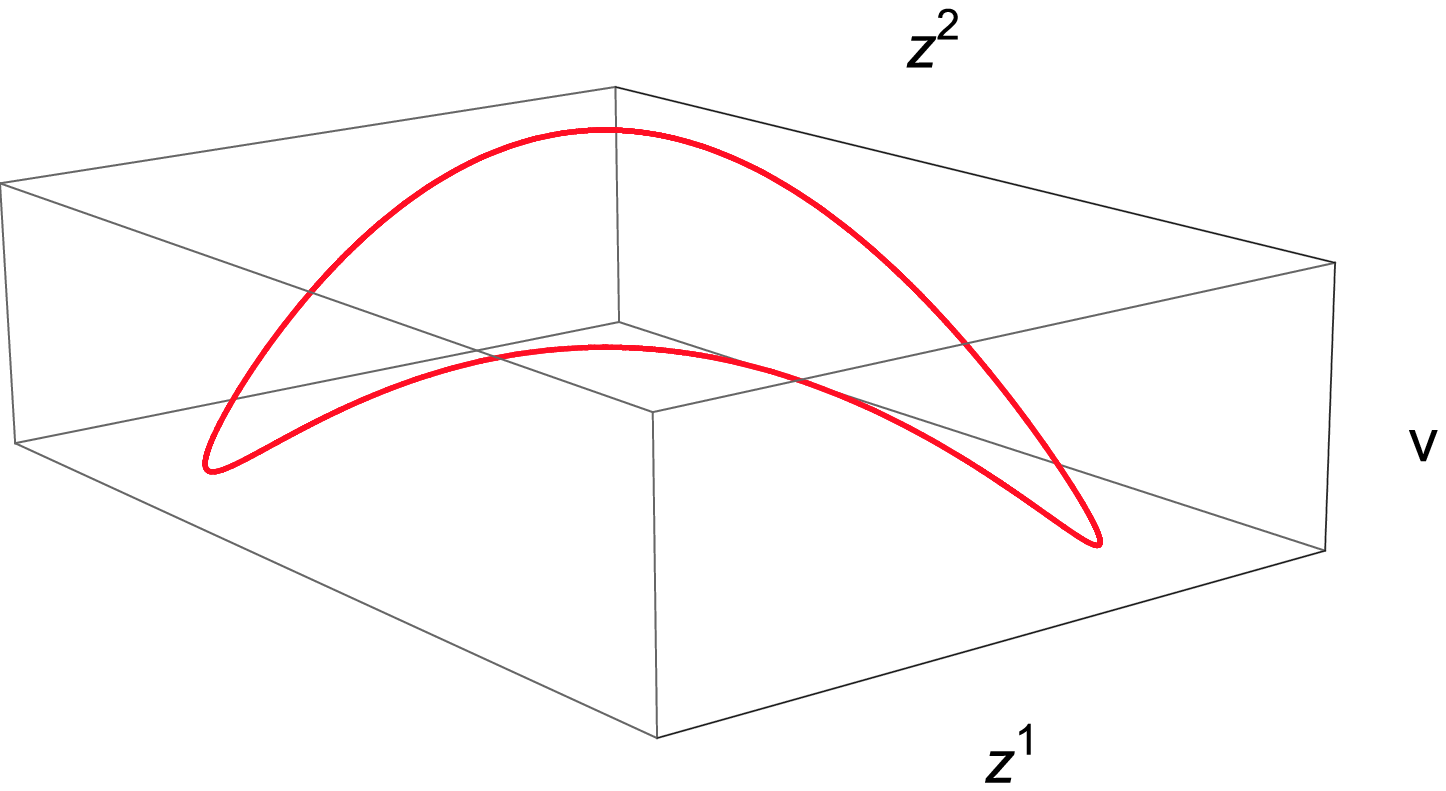}
\caption{The left-hand plot highlights (in red) an element of the twisted null congruence propagating in the
null $u$-direction. The right-hand plot illustrates the periodicity in $u$ of the geodesic, which 
therefore appears closed when plotted in $z^1, z^2, v$ coordinates.}
\label{GeodesicPlots}
\end{figure}

To visualise these geodesics, it is sufficient to select an individual element of the congruence
by choosing values for the Rosen coordinates $V, X^1, X^2$. The corresponding curves 
(with $V=0$, $X^1=1$ and $X^2=0$ in the Ricci-flat OS metric $a=1$, $b=-1$) 
are plotted in Figure \ref{GeodesicPlots}. 
In the transverse space, as the geodesic progresses along $u$, the 
coordinates $z^1, z^2$ describe an ellipse with period $u=2\pi/\l$. Meanwhile, the null coordinate
$v$ is oscillating sinusoidally with half the period. The full geodesic is therefore periodic in $u$
with period $2\pi/\l$, as can be seen in the right-hand figure.

Evidently, this geodesic is part of a twisted null congruence, with $\omega_{ij} = \L \e_{ij}$
and $\L = -1/2$.
There is no focusing, and there are no conjugate points, consistent with the vanishing of the expansion 
optical scalar $\theta$ for the congruence. The shear $\s_{ij}$ is apparent in the different 
amplitudes for the oscillations in $z^1$ and $z^2$ for $\b \neq 1$. (Recall from (\ref{d23}) that
the shear is non-vanishing for a profile function $h_{ij}$ with $a\neq b$, which from (\ref{d15})
implies $\b \neq 1$.)

The absence of conjugate points is worth noting. Generically, null congruences in a plane wave
spacetime will focus to conjugate points provided the Ricci tensor satisfies the null energy
condition ($R_{uu} \ge 0$ in our conventions) by virtue of the negativity of the r.h.s.~of the
Raychaudhuri equation (\ref{b12}) for the expansion scalar $\theta$. These conjugate points 
played a key r\^ole in our work on quantum loop effects in wave propagation in curved spacetime,
where they are associated with singularities in the relevant Green functions 
and determine key features of the refractive index in the quantum field theory 
\cite{Hollowood:2007ku, Hollowood:2008kq}.
However, this only holds in the absence of twist. Since $\omega_{ij} = \L \e_{ij}$, the term
$\tr\, \omega^2$ in (\ref{b12}) is {\it negative}.\footnote{Note that we use the notation $(\omega^2)_{ij}
= \omega_{ik}\omega^k{}_j$, which introduces a minus sign in $\tr\, \omega^2$ relative to a 
common convention for the Raychaudhuri equations (see {\it e.g.}~\cite{Chandrasekhar:1985kt}.}
So even for non-Ricci flat spacetimes, expansion-free null congruences can be supported by
balancing the twist contribution to (\ref{b12}) against the shear and Ricci terms.

\subsection{Newman-Penrose tetrad in co-rotating coordinates}

As we have seen, the Newman-Penrose tetrad associated with a null geodesic is a powerful tool for
analysing the geometry of gravitational plane waves. Here, we find the Newman-Penrose basis for the 
null geodesic congruence as described above in the co-rotating coordinates and use this to find the
corresponding Penrose limit. Naturally this should reproduce the result of section \ref{sect2.4},
though the construction involves some interesting subtleties which were observed in the original
description \cite{OS} of the Ozsv{\'a}th-Sch{\"u}cking spacetime.

A short calculation shows that the tetrad\footnote{The corresponding covectors are
\begin{align*}
L_\m &= \left(\dot{v} + z(h_0 + {\bf 1})z + z^2 \dot{z}^1 - z^1 \dot{z}^2 , ~1, ~z^2 + \dot{z}^1, 
~-z^1 + \dot{z}^2 \right) \ , \nonumber \\
N_\m &= \left(-1,~0,~0,~0\right) \ , \hskip1cm
a_\m = \left(-\dot{z}^1,~0,~1,~0\right) \ , \hskip1cm
b_\m = \left(-\dot{z}^2,~0,~0,~1\right) \ .
\end{align*}}
\begin{equation}
L^\m = k^\m = \begin{pmatrix}1\\ \dot{v} \\ \dot{z}^1 \\ \dot{z}^2\end{pmatrix} \ , \hskip0.8cm
N^\m = \begin{pmatrix}0 \\ -1 \\ 0 \\ 0\end{pmatrix} \ , \hskip0.8cm
a^\m = \begin{pmatrix}0 \\ -\dot{z}^1 - z^2 \\ 1\\ 0\end{pmatrix} \ , \hskip0.8cm
b^\m = \begin{pmatrix}0 \\ -\dot{z}^2 + z^1 \\ 0 \\ 1\end{pmatrix}  \hskip0.8cm
\label{d32}
\end{equation}
with $m^\m = \tfrac{1}{\sqrt{2}}\left(a^\m + i b^\m\right)$, satisfies the required Newman-Penrose 
conditions $L^2 = N^2 = m^2 = 0$, $L.N=-1$, $m.\bar{m} = 1$, {\it etc}. The first derivatives 
$\dot{v}, \dot{z}^1, \dot{z}^2$ can be eliminated in favour of $z^1, z^2$ immediately using
(\ref{d27}) and (\ref{d29}).

However, we still need to check that this tetrad is parallel-transported along the geodesic $\c$. 
We do indeed find $L^\m D_\m L^\n = 0$ and $L^\m D_\m N^\n$, the former especially 
requiring care.\footnote{For example, using the 
non-vanishing Christoffel symbols in footnote \ref{fn2}, we find for the $L^\n$ component,
\begin{align*}
L^\m D_\m L^\n &= \dot{z}^i \partial_i L^\n + \Gamma^v_{uu} + 2 \Gamma^v_{ui}\dot{z}^i \nonumber \\
&= z^1 z^2 \left(a - b + 2a \frac{\l}{\b} - 2b \b\l + 2 \frac{\l^3}{\b} - 2 \b \l^3 \right) = 0 \ ,
\end{align*}
after using (\ref{d15}) to simplify the terms involving $\l^3$.}
This is not true, however,  for the transverse vectors, where we find
\begin{equation}
L^\m D_\m a^\n = \begin{pmatrix} 0 \\-(1+\l\b)z^1 \\0 \\-1 \end{pmatrix} \ , \hskip2cm
L^\m D_\m b^\n = \begin{pmatrix} 0 \\-(1+\l/\b)z^1 \\1 \\0 \end{pmatrix} \ .
\label{d33}
\end{equation}
The resolution is to define new transverse vectors
\begin{equation}
\begin{pmatrix}A^\m \\B^\m\end{pmatrix} = O(u) \begin{pmatrix} a^\m \\b^\m \end{pmatrix} \ ,
\label{d34}
\end{equation}
for which we find
\begin{align}
L^\m D_\m A^\n &= \cos u \left(L^\m D_\m a^\n + b^\n\right) + \sin u \left(L^\m D_\m b^\n - a^\n\right) 
= 0 \ , \nonumber \\
L^\m D_\m B^\n &= \cos u \left(L^\m D_\m b^\n - a^\n\right) - \sin u \left(L^\m D_\m a^\n + b^\n\right) 
= 0 \ ,
\label{d35}
\end{align}
since the bracketed contributions vanish by comparing (\ref{d31}) and (\ref{d32}) using (\ref{d27}).
The full set of basis vectors $L^\m, N^\m$ and $M^\m = \tfrac{1}{\sqrt2}\left(A^\m + B^\m\right)$ 
now satisfies all the required properties for a Newman-Penrose tetrad parallel-transported along $\c$.

What this shows is that even working in the stationary coordinate system $(u,v,z^1,z^2)$ for the
generalised OS metric, requiring that the transverse Newman-Penrose vectors are parallel-transported 
along the null geodesic {\it reintroduces} the rotation $O(u)$ which is manifest in the Brinkmann 
coordinate description of the metric.

This also resolves what at first sight seems mysterious in deriving the Penrose limit associated with 
a null geodesic in the generalised OS metric described in the stationary coordinates,
{\it viz.} how does the Penrose limit reproduce the homogeneous plane wave metric including the
rotation factor $O(u)$ as shown in section \ref{sect2.4}. In the Newman-Penros formalism, the resolution
is especially elegant. Writing (\ref{b29}) in terms of $A^\m, B^\m$ rather than $M^\m, \bar{M}^\m$, we 
have the Penrose limit profile function $\hat{h}_{ij}$ in the form,
\begin{align}
\hat{h}_{ij} &= - \begin{pmatrix} C_{LALA} + \tfrac{1}{2}R_{LL} &\,\, C_{LALB} \\
{}&{} \\
C_{LBLA} &\,\, C_{LBLB} + \tfrac{1}{2} R_{LL} \end{pmatrix} \nonumber \\
{}&{} \nonumber \\
&= - O(u)\,\begin{pmatrix} C_{LaLa} + \tfrac{1}{2}R_{LL} &\,\, C_{LaLb} \\
{}&{} \\
C_{LbLa} &\,\, C_{LbLb} + \tfrac{1}{2} R_{LL} \end{pmatrix}\, O^T(u) \ .
\label{d36}
\end{align}
Evaluating the Ricci and Weyl tensor components using (\ref{d5a}), we then quickly find
\begin{equation}
\hat{h}_{ij} ~=~ O(u) \,\begin{pmatrix} a &\,\,0\\ 0 &\,\, b\end{pmatrix}\, O^T(u) ~=~ h_{ij} \ .
\label{d37}
\end{equation}
That is, the correct choice of parallel-transported Newman-Penrose tetrad automatically reinstates the 
implicit rotation in the Brinkmann coordinate description of the homogeneous plane wave. 
This confirms in this explicit example of geodesics belonging to a twisted null congruence in the
generalised OS spacetime that the associated Penrose limit simply reproduces the original
homogeneous plane wave metric.

\subsection{Isometries}\label{sect 4.6}

In section \ref{sect 3}, we described the extended isometry algebra for a homogeneous plane
wave, with the generators $Q_r$ and $P_r$ defined in (\ref{c9}) with oscillator solutions
$f^i_{(r)}$ and $g^i_{(r)}$ satisfying canonical boundary conditions (\ref{b20}).
We are free, however, to choose any independent linear combination of these to define
$Q_r$, $P_r$ and such a redefinition will of course change the standard form of the
algebra (\ref{c5}), (\ref{c6}).

As we saw when describing the isometries in terms of Rosen coordinates, a
particularly natural choice is to define the generator $Q_r$ with an oscillator solution 
$F^i{}_{(r)} = E^i{}_a \d^a{}_r$, since this reflects the nature of the twisted congruence.
For $P_r$, we require a second, independent solution $G^i{}_{(r)}$. 
As we now show, an interesting choice in this model is to take $G = \dot{E} - \epsilon E$.

By definition then, from (\ref{c17}), (\ref{c18}) we immediately have
\begin{equation}
\left[X, Q_r\right] = P_r \ ,
\label{d38}
\end{equation}
for the commutator with the extra generator $X$ related to $u$-translations. 
To find the commutator of $X$ with $P_r$, we need to iterate this construction
and determine the combination $\dot{G} - \epsilon G$. Considering the 
generalised OS model and using the explicit expressions for $E$, $\Omega$ and $\omega$
from section \ref{sect 4.3}, we can easily show,
\begin{align}
\dot{G} - \epsilon G &= \left(h - 1 - 2 \epsilon \Omega\right) E \nonumber \\
&= O(u) \begin{pmatrix} a+1 + 2\lambda\b &\,\, 0 \\
0&\,\, \b\left(b + 1 + 2 \lambda/\b\right) \end{pmatrix} O(\lambda u)  \nonumber \\
&= - \lambda^2 E \ ,
\label{d39}
\end{align}
using the usual equations (\ref{d15}) for $\lambda, \b$. It follows directly that
\begin{equation}
\left[X, P_r\right] = - \lambda^2\,Q_r \ .
\label{d40}
\end{equation}

For the remaining commutators involving $Q_r$ and $P_r$, we need to evaluate the 
relevant Wronskians. A short calculation using the zweibein of section \ref{sect 4.3}
gives first $G = \lambda E \epsilon$ and then,
\begin{equation}
W(F,F) = 2\,\bomega \ , ~~~~~~~~ W(F,G) = 2\,\lambda\bomega \epsilon \ ,
~~~~~~~~ W(G,G) = 2\, \lambda^2 \bomega \ .
\label{d41}
\end{equation}
The corresponding commutators are (compare (\ref{c16})),
\begin{align}
\left[Q_r,Q_s\right] = &- W_{rs}(F,F) \,Z \ , ~~~~~~~~ 
\left[P_r,P_s\right] = - W_{rs}(G,G) \, Z \ , \nonumber \\
{} \nonumber \\
&\left[Q_r,P_s\right] = - W_{rs}(F,G) \,Z \ .
\label{d42}
\end{align}
In particular, this shows how the twist enters into the non-vanishing commutators
of $Q_r$ (and $P_r$) with itself.  An alternative presentation using the model-dependent
constants $\lambda, \b$ and $\Lambda$ defined in section \ref{sect 4.3} 
(where $\b = \pm\sqrt{2}, \, \lambda = \mp\sqrt{2},\,\Lambda = -1/2$ for the Ricci-flat 
OS metric) is then
\begin{align}
\left[Q_r,Q_s\right] = \,\,&2\b\Lambda\,\epsilon_{rs}\,Z \ , ~~~~~~~~~~~~
\left[P_r,P_s\right] = 2 \lambda^2 \b \Lambda\,\epsilon_{rs}\,Z \ , \nonumber \\
{} \nonumber \\ 
&\left[Q_r,P_s\right] = - 2\lambda \b \Lambda \,\d_{rs}\,Z \ .
\label{d43}
\end{align}

\end{document}